\documentclass[twocolumn,showpacs,preprintnumbers,graphicx,amsmath,amssymb,reprint,floatfix]{revtex4-1}
\setcitestyle{super}
\pdfoutput=1
\usepackage{graphicx}
\usepackage{dcolumn}
\usepackage{bm}
\usepackage{times}
\usepackage{multirow}
\usepackage{longtable}
\usepackage{natbib,hyperref}
\usepackage{color}

\begin{document}


\title{Mapping the band structure of GeSbTe phase change alloys around the Fermi level}
\author{J. Kellner$^1$}

\author{G. Bihlmayer$^2$}
\author{M. Liebmann$^1$}
\author{S. Otto$^3$}
\author{C. Pauly$^1$}
\author{J. E. Boschker$^4$}
\author{V. Bragaglia$^4$}
\author{S. Cecchi$^4$}
\author{R. N. Wang$^4$}
\author{V. L. Deringer$^5$}
\thanks{Present address: Engineering Laboratory, University of Cambridge, Cambridge CB2 1PZ, United Kingdom}
\author{P. K{\"u}ppers$^1$}
\author{P. Bhaskar$^1$}
\author{E. Golias$^6$}
\author{J. S\'{a}nchez-Barriga$^6$}
\author{R. Dronskowski$^5$}
\author{T. Fauster$^3$}
\author{O. Rader$^6$}
\author{R. Calarco$^4$}
\author{M. Morgenstern$^1$}
\email[] {mmorgens@physik.rwth-aachen.de}
\affiliation{ $^1$II. Physikalisches Institut B and JARA-FIT, RWTH Aachen University, D-52074 Aachen, Germany}
\affiliation{ $^2$Peter Gr{\"u}nberg Institute (PGI-1), Forschungszentrum J{\"u}lich, D-52428 J{\"u}lich, Germany}
\affiliation{ $^3$Lehrstuhl f{\"u}r Festk{\"o}rperphysik, Universit{\"a}t Erlangen-N{\"u}rnberg, D-91058 Erlangen, Germany}
\affiliation{ $^4$Paul-Drude-Institut f{\"u}r Festk{\"o}rperelektronik Berlin, D-10117 Berlin, Germany}
\affiliation{ $^5$Institute of Inorganic Chemistry, RWTH Aachen University, D-52074 Aachen, Germany}
\affiliation{ $^6$Helmholtz-Zentrum f{\"u}r Materialien und Energie BESSY II, D-14109 Berlin, Germany}

\date{\today}

\begin{abstract}
Phase change alloys are used for non-volatile random access memories exploiting the conductivity contrast between amorphous and metastable, crystalline phase. However, this contrast has never been directly related to the electronic band structure. Here, we employ photoelectron spectroscopy to map the relevant bands for metastable, epitaxial GeSbTe films. The constant energy surfaces of the valence band close to the Fermi level are hexagonal tubes with little dispersion perpendicular to the (111) surface. The electron density responsible for transport belongs to the tails of this bulk valence band, which is broadened by disorder, i.e., the Fermi level is 100\,meV above the valence band maximum. This result is consistent with transport data of such films in terms of charge carrier density and scattering time. In addition, we find a state in the bulk band gap with linear dispersion, which might be of topological origin.
\end{abstract}


\maketitle 

\section*{Introduction}

Phase change alloys are the essential components for optical data storage (DVD-RW, Blu-ray Disc) and for electrically addressable phase-change random-access memories (PC-RAM)\cite{Wuttig2007,Wuttig2012}. The latter are envisioned to become more energy efficient using interfacial phase-change memories, whose phase change has been related to a topological phase transition\cite{Tominaga2014}. Phase change alloys are typically chalcogenides consisting of Ge, Sb and Te (GST) with Ge$_2$Sb$_2$Te$_5$ (GST-225) being the prototype\cite{Wuttig2007,Deringer2015}. They exhibit three different structural phases: an amorphous, a metastable rock salt, and a stable trigonal phase. Switching the system from amorphous to metastable leads to a large contrast in electrical conductivity and optical reflectivity, which is exploited for data storage\cite{Ovshinsky1968,Yamada1987}. Such switching favorably occurs within nanoseconds\cite{Yamada1991,Loke2012} and at an energy cost down to 1\,fJ for a single cell\cite{Xiong2011}.

\begin{figure}
\includegraphics[width=1\linewidth]{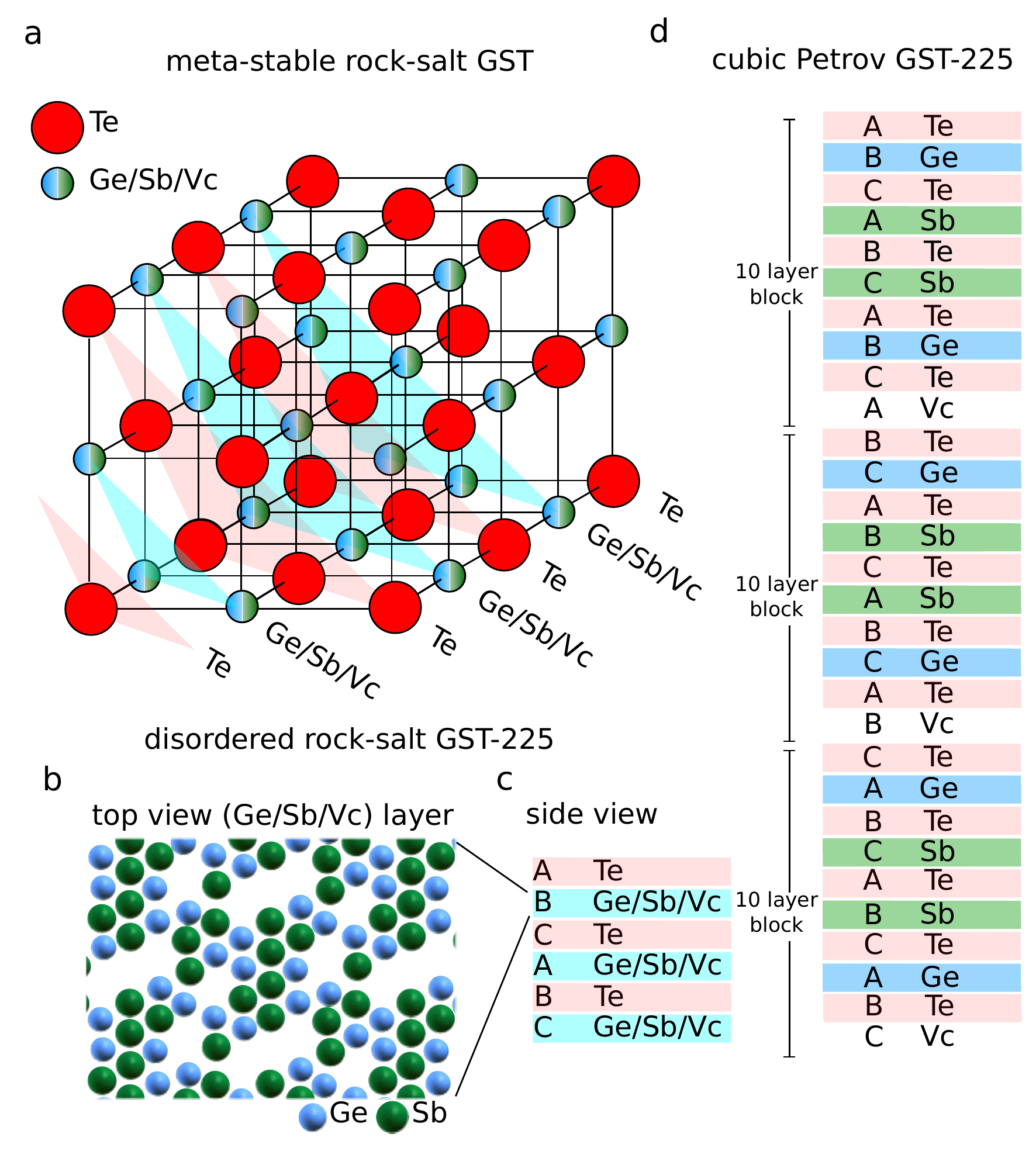}
\caption{\label{fig:Structure} Structural models of Ge$_2$Sb$_2$Te$_5$ (GST-225): (a) metastable rock salt structure: red circles: Te, striped circles: Ge, Sb or vacancy (Vc); (111) planes are marked with alternating colors (red: Te, turquoise: Ge/Sb/Vc); (b) DFT-optimized structure of the disordered subsurface layer of a GST-225 slab, blue: Ge, green: Sb; (c) layer structure of one unit cell of metastable rock salt GST-225 exhibiting ABC stacking; the $2\times3$ layers in the unit cell result from the two chemically distinct layers and the three distinct stacking positions; (d) same as (c), but for the cubic Petrov-phase of GST-225 exhibiting a unit cell of $10\times3$ layers with 10 layers due to the alternating chemistry and a 3-fold repetition due to the ABC stacking.}
\end{figure}

The technologically relevant, metastable phase\cite{Park2007}, usually obtained by rapid quenching from the melt, has a rock salt like structure with Te atoms at one sublattice and a mixture of randomly distributed Ge, Sb and vacancies (Vcs) on the other sublattice (Fig.\ \ref{fig:Structure}a, b)\cite{Matsunaga2004,Matsunaga2006,Matsunaga2008,Silva2008}. The stable phase consists of hexagonally close-packed layers of either Ge, Sb or Vcs with hexagonal layers of Te in between. Hence, the Vc layers bridge adjacent Te layers \cite{Petrov1968,Kooi2002}.  The stable phase has trigonal symmetry and is distinct in stacking of the hexagonal layers from the regular ABC stacking within the rock salt like metastable phase (Fig.\ \ref{fig:Structure}c, d).

In the metastable phase, the disorder on the (Ge,Sb,Vc) site leads to Anderson localization of the electrons \cite{Siegrist2011}.  The localization is lifted by annealing due to the respective continuous ordering of the Ge, Sb, and Vcs into different layers \cite{Schneider2012,Zhang2012,Bragaglia2016}. This is accompanied by a shift of the Fermi level $E_{\rm F}$ towards the valence band (VB)\cite{Siegrist2011,Subramaniam2009}. However, the corresponding Fermi surface is not known as well as the exact position of $E_{\rm F}$, such that it is difficult to understand the electrical conductivity in detail.

Most of the electrical transport measurements so far were conducted using polycrystalline GST \cite{Siegrist2011,Volker}, such that many established tools requiring crystallinity of the samples could not be applied. Only recently, epitaxial films of single crystalline quality have been achieved by molecular beam epitaxy (MBE)\cite{Katmis2011,Rodenbach2012,Bragaglia2014,Bragaglia2016,Cecchi2017}. These films have been probed so far by X-ray diffraction (XRD), electron microscopy\cite{Katmis2011,Rodenbach2012,Bragaglia2014,Bragaglia2016,Mitrofanov2016,Cecchi2017}, magnetotransport studies\cite{Bragaglia2016}, Raman spectroscopy, and Fourier transform infrared spectroscopy\cite{Cecchi2017, Bragaglia2016a}. Most importantly, it was found that the epitaxial GST films are in the technologically relevant rock salt phase, but often exhibit ordering of the vacancies in separate layers\cite{Bragaglia2016}.

Here, we provide the first detailed measurement of the band structure of such epitaxial films
by angular resolved photoelectron spectroscopy (ARPES). We focus on the nominal composition GST-225, and employ an ultrahigh-vacuum (UHV) transfer from the MBE system to prevent surface oxidation\cite{Yashina2008} (see methods).
Within the whole Brillouin zone (BZ), we find an M-shaped bulk VB in all directions parallel to the surface. This is in qualitative agreement with density functional theory (DFT) calculations of the cubic adaption of the trigonal Petrov phase\cite{Petrov1968,Sun2006}, sketched in Fig.\ \ref{fig:Structure}d. For brevity, we call this structure the cubic Petrov phase. Connecting the VB maxima of the experimental data results in a hexagonal tube at an energy about 100\,meV below $E_{\rm F}$.  Hence, the classical Fermi volume of a strictly periodic system would be zero, which contradicts the observation of metallic conductivity\cite{Bragaglia2016}. This apparent contradiction is solved by the significant broadening of the $E(\mathbf{k})$ states due to disorder, such that there is still considerable weight of the valence band states above $E_{\rm F}$. The sum of these weights results in a charge carrier density $n_{\rm eff}$ consistent with the charge carrier density obtained from Hall measurements of the MBE films. The width of the states is, moreover, compatible with the scattering time deduced from the transport data.
Such a detailed description of electrical transport provides a significant improvement over more simplistic models based on a parabolic and isotropic valence band as used so far\cite{Siegrist2011,Volker}.

Additionally, we find an electronic band within the fundamental bulk band gap of the metastable phase by two-photon ARPES. This band exhibits a largely isotropic, linear dispersion and circular dichroism such as known for topological surface states (TSS) \cite{Wang2013,Scholz2013,Wang2013,Seibel2015}. We also find states close to the VB maximum with a strong in-plane spin polarization perpendicular to $\mathbf{k}$ by conventional ARPES again similar to TSSs. A non-trivial topology of GST-225 has indeed been predicted for certain stacking configurations by DFT calculations\cite{Kim2010,Sa2011,Kim2012,Sa2012,Silkin2013} and has been conjectured from the M-type VB dispersion\cite{Pauly2013}.
Assuming that  the Dirac-type state is indeed a TSS and, hence, cuts $E_{\rm F}$, it would contribute to the electronic transport. It would even dominate the conductivity,
if its mobility $\mu$ is larger than $0.1\,\mathrm{m}^2(\mathrm{Vs})^{-1}$. This is lower than the best TSS mobilities found in other topological insulators such as Bi$_2$Se$_3$ and  BiSbTeSe$_2$ films ($\mu \simeq 1\,\mathrm{m}^2(\mathrm{Vs})^{-1}$)\cite{Oh2015,Xu2016}.

\section*{Results}
\subsection*{Constant energy surfaces}

\begin{figure*}
\includegraphics[width=\textwidth]{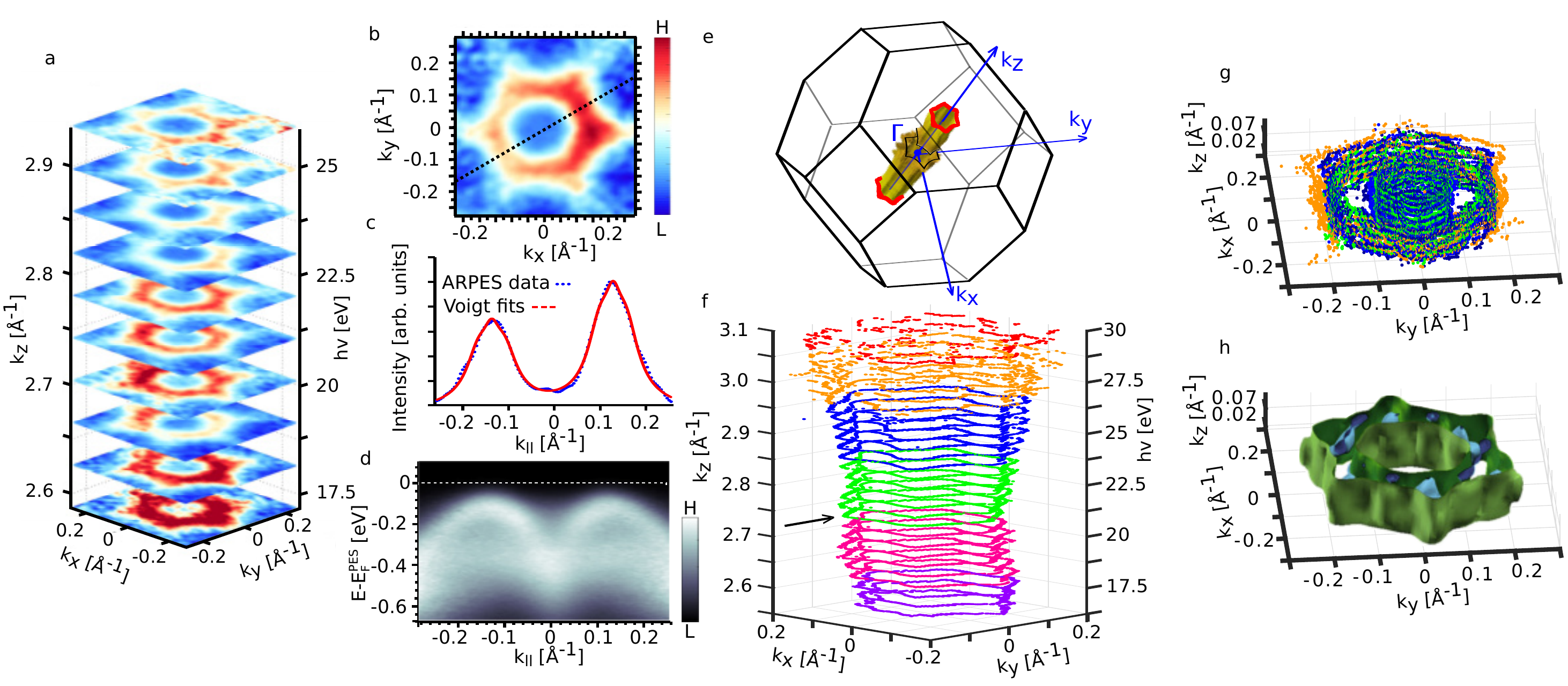}
\caption{\label{fig:ARPES_MDC} ARPES and DFT data for different photon energies $h\nu$:
(a) constant  energy cuts (CECs) at the Fermi level $E_{\rm F}^{\rm PES}$ for different $h\nu$. We deduce $k_z$ as marked on the left from $h\nu$ using an inner potential $E_{\rm inner} =14$\,eV (methods);
(b) CEC at $E_{\rm F}^{\rm PES}$ for $k_{z}=2.83$\,\AA$^{-1}$  with direction equivalent to $k_y$ marked by a dotted black line;
(c) momentum distribution curve (MDC) along dotted black line in (b); two Voigt peaks (red line) are fitted to the ARPES data (blue dots);
(d) Energy momentum cut (EMC) along the dotted line marked in (b) at $k_{z}=2.83$\,\AA$^{-1}$, dashed line marks $E_{\rm F}^{\rm PES}$;
(e) Brillouin zone (BZ) of metastable, disordered rock salt GST-225 (structure in Fig.\ \ref{fig:Structure}a) with principal $\mathbf{k}$ directions marked; the measured pseudo Fermi surface is shown in gold including the missing part due to vanishing photoelectron intensity (methods); the Fermi lines cutting the BZ side planes in (111) direction are drawn in red;
(f) Voigt peak positions at $E_{\rm F}^{\rm PES}$ from MDC fits as in (c); colors are alternating in $\Delta k_z = 0.12$\,\AA$^{-1}$, being the $k_z$ size of the BZ of the cubic Petrov phase as used in the DFT calculations; the tentative BZ boundary of the metastable rock salt phase at $k_{z}=2.72$\,\AA$^{-1}$ is marked by a black arrow; it is used to determine $E_{\rm inner}$; the additionally given photon energies (right scale) correspond to the average diameter of the hexagons;
(g) $E_{\rm peak, 1}(\mathbf{k})$ at $E-E_{\rm F}^{\rm PES}=-200$\,meV after back-folding along $k_z$ into the BZ of the cubic Petrov phase (same color code as in (f));
(h) constant energy surfaces (CESs) of the valence bands (VBs) of the cubic Petrov phase according to DFT at the experimental $E-E_{\rm F}^{\rm PES}=-200$\,meV, i.e., after shifting $E_{\rm F}^{\rm DFT}$ by $100$\,meV upwards, such that the VB maxima in DFT and ARPES match.}
\end{figure*}

ARPES experiments were performed at 29 different photon energies $h\nu=16-31$\,eV with a step size of $0.5$\,eV. This allows a detailed determination of the $k_{z}$ dispersion of the bands ($k_z$: wave vector perpendicular to the surface). Using the estimated crystal potential $E_{\rm inner} = 14$\,eV (methods), the chosen $h\nu$ relate to $k_{z}=2.55-3.09$\,\AA$^{-1}$. The ARPES spectra show an inverted M-shaped VB in energy-momentum cuts (EMCs) taken along the surface plane (Fig.\ \ref{fig:ARPES_MDC}d). The independently measured Fermi level $E_{\rm F}^{\rm PES}$ is well above the VB maximum. Both is in line with earlier, less extensive results\cite{Pauly2013}. We label $E_{\rm F}^{\rm PES}$ additionally with the superscript PES, since it differs from $E_{\rm F}^{\rm DFT}$ in DFT calculations with respect to the VB maximum.  Figure\ \ref{fig:ARPES_MDC}a displays constant energy cuts (CECs) of the normalized photoelectron intensity (methods) at $E_{\rm F}^{\rm PES}$ for selected $k_{z}$. To determine peak positions, momentum distribution curves (MDCs) are extracted and fitted by Voigt peaks (Fig.\ \ref{fig:ARPES_MDC}b,c). The resulting peak momenta $\mathbf{k}$ form a hexagonal tube (Fig.\ \ref{fig:ARPES_MDC}e,f) called the {\it pseudo} Fermi surface of GST-225. We call it {\it pseudo}, since the peak energies $E_{\rm peak}(\mathbf{k})$ resulting from fits of energy distribution curves (EDCs) do not cross $E_{\rm F}^{\rm PES}$ for any $\mathbf{k}$,
as visible, e.g., in Fig.\ \ref{fig:ARPES_MDC}d. Consequently, there are no band centers at $E_{\rm F}^{\rm PES}$ as required for a conventional Fermi surface \cite{Medjanik2017}. Only the tails of the broadened energy peaks cross  $E_{\rm F}$. The sizes of the hexagons of the pseudo Fermi surface slightly vary with $h\nu$, i.e. along $k_z$, with minimal diameter at $h\nu=21$\,eV (arrow in Fig.\ \ref{fig:ARPES_MDC}f). We conjecture (in accordance with DFT) that this minimum corresponds to the BZ boundary and, hence, use it to determine $E_{\rm inner}=14$\,eV, unambiguously relating $h\nu$ to $k_z$ (methods).
For the Fermi wave number in $x$ ($y$) direction, we find $k_{{\rm F},x}= 1.52\pm 0.3\,\mathrm{nm}^{-1}$ ($k_{{\rm F},y}= 1.43\pm 0.2\,\mathrm{nm}^{-1}$), where the $\pm$ interval describes the full variation along $k_z$.
Hence, with a precision of 20\,\%, the pseudo Fermi surface is a hexagonal tube without dispersion along $k_z$.

The EDCs consist of up to two peaks down to $E-E_{\rm F}^{\rm PES}=-1$\,eV for all probed $\mathbf{k}$. These peaks are fitted by two Voigt peaks with peak centers $E_{{\rm peak}, j}(\mathbf{k})$ ($j=1,2$). The highest peak energy
for all $\mathbf{k}$, i.e., the VB maximum, is found at $E_{{\rm peak},1}(\mathbf{k})-E_{\rm F}^{\rm PES}= -105\pm 10$\,meV with $\mathbf{k}=(0\pm 0.02, 1.53\pm 0.02, 25.8\pm 0.2)\,\mathrm{nm}^{-1}$, as well as at equivalent $\mathbf{k}$ points. Projecting back to the first BZ, we get $\mathbf{k}=(0, 1.53, 1.6)\,\mathrm{nm}^{-1}$, i.e., the VB maximum is offset from $\Gamma$ also in $k_z$ direction. 

\begin{figure*}
\includegraphics[width=\textwidth]{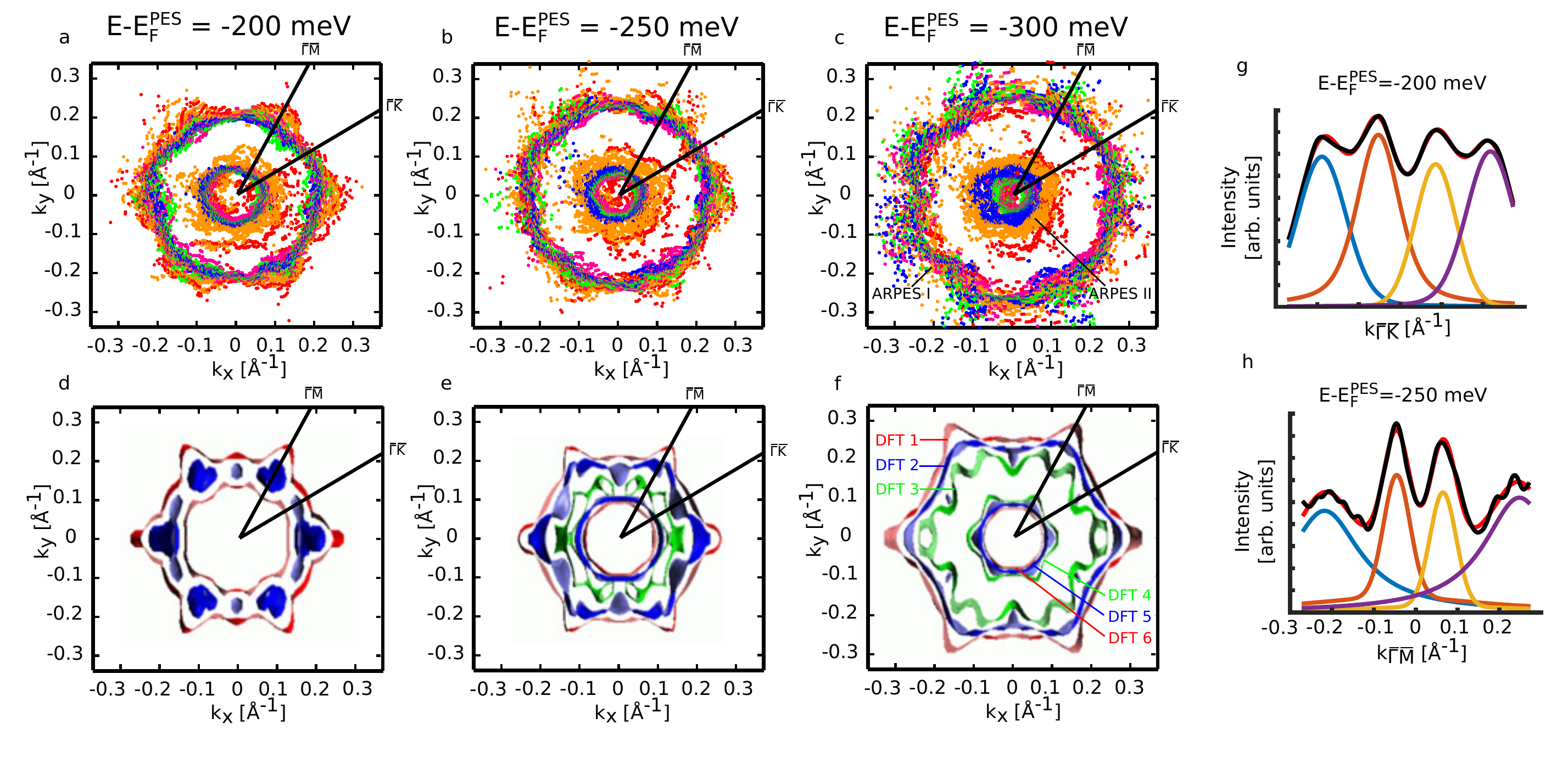}
\caption{\label{fig:ARPES_DFT_MDC}Comparison between ARPES and DFT at different binding energies $E-E_{\rm F}^{\rm PES}$:
(a)-(c) Voigt peak positions $E_{\rm peak, 1}(\mathbf{k})$, deduced from MDC fittings, projected to the ($k_x$, $k_y$) plane; the marked $\overline{\Gamma} \overline{M}$ and $\overline{\Gamma} \overline{K}$  belong to a surface BZ projection; the color code is the same as in Fig.\ \ref{fig:ARPES_MDC}(f)-(g);
(d)-(f) corresponding CESs from DFT calculations exhibiting three Sb p-type VBs (different colors) within the BZ of the cubic Petrov phase; numberings in (c) and (f) label the different bands;
(g)-(h) measured MDCs (black curves) along $\mathbf{k}_\parallel$ directions and at $E-E_{\rm F}^{\rm PES}$ as marked. Red fit curves consist of the displayed four Voigt peaks (blue, orange, yellow and violet curve).}
\end{figure*}

Constant energy surfaces (CESs) of $E_{{\rm peak}, j}(\mathbf{k})$ are constructed below the VB maximum \cite{Huefner,Matsdorf1998,Zhang2011,Medjanik2017}.
They are compared with CESs from DFT calculations, which require periodic boundary conditions, i.e., a distinct order within the Ge/Sb/Vc layer. We have chosen the cubic Petrov phase (Fig.\ \ref{fig:Structure}d) to represent the metastable ABC stacking of the rock salt structure employing chemically pure Sb, Ge and Vc layers \cite{Sun2006,Pauly2013}. Since the corresponding DFT BZ is reduced in $k_{z}$ direction by a factor of $5$ with respect to the disordered rock salt phase (Fig.\ \ref{fig:Structure}a-c), the ARPES data have to be back-folded into a $k_z$ range of $\Delta k_{z}=0.12$\,\AA$^{-1}$ for comparison. Therefore, the measured $k_{z}$ data are divided into parts covering $\Delta k_{z}=0.12$\,\AA$^{-1}$ each (see color code in Fig.\ \ref{fig:ARPES_MDC}f) and projected accordingly. Results at $E-E_{\rm F}^{\rm PES}=-200$\,meV are shown in Fig.\ \ref{fig:ARPES_MDC}g, where each MDC has been fitted by four Voigt peaks as exemplary shown in Fig.\ \ref{fig:ARPES_DFT_MDC}g. The qualitative agreement with the DFT CESs (Fig.\ \ref{fig:ARPES_MDC}h) is reasonable, in particular, for the outer hexagon. Such agreement is also found at other energies as shown in Fig.\ \ref{fig:ARPES_DFT_MDC}a-f, where the different $k_z$ values are projected to the $(k_x,k_y)$ plane. However, quantitative differences are apparent as discussed in Supplementary Note 1.

\subsection*{Effective charge carrier density from ARPES and magnetotransport}  \label{CCD}

\begin{figure*}
\includegraphics[width=\textwidth]{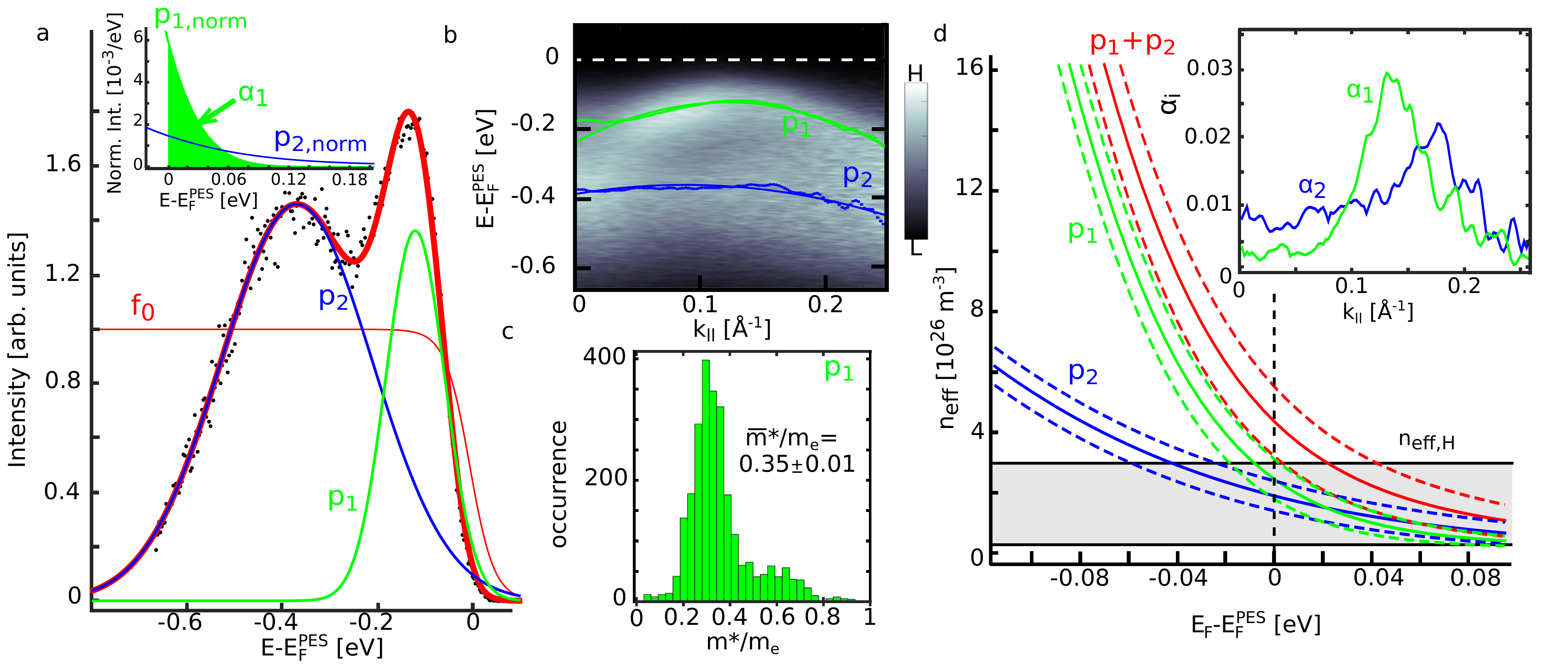}
\caption{\label{fig:ARPES_EDC} Charge carrier density $n_{\rm eff}$ and curvature parameter $m^\star$ deduced from ARPES:
(a) EDC at $\mathbf{k}=(0.0,0.13,2.73)$\,\AA$^{-1}$ (black dots) with fit curve (red) consisting of two Voigt peaks ($p_1$ = green, $p_2$ = blue) multiplied by the Fermi distribution function $f_0(E, T=300\,\mathrm{K})$ (thin red line);
inset: zoom around $E_{\rm F}^{\rm PES}$ displaying peak tails, after scaling each peak area to unity; only the colored tail of peak 1 (marked $\alpha_1$) contributes to $n_{\rm eff,1}$;
(b) EMC at $k_{\rm z}=2.73$\,\AA$^{-1}$ with marked peak positions $p_1$ (green dots) and $p_2$ (blue dots) resulting from fits of EDCs as in (a); parabolic fits to these points (accordingly colored full lines), used to determine $m^\star$, are added;
(c) histogram of resulting $m^\star / m_{\rm e}$ belonging to the band of $p_1$ using $90$ different  azimuths in ($k_x$, $k_y$) direction for $29$ different $k_z$ values; errors for individual $m^\star / m_{\rm e}$ are about the bin width;
(d) red full line: $n_{\rm eff}$ as calculated from the ARPES data according to Eq.\ \ref{eq:alpha} for a hypothetically varying $E_{\rm F}$ with respect to the measured $E_{\rm F}^{\rm PES}$; different contributions from $p_1$ (green full line) and $p_2$ (blue full line) are labeled;  error intervals are depicted by correspondingly colored dashed lines; charge carrier densities from Hall measurements ($n_{\rm eff, H}$) of identically prepared GST films are added as a grey box;
inset: $\alpha_j(\mathbf{k})$ (relative part of the Voigt peak $p_j$ at $\mathbf{k}$ above $E_{\rm F}^{\rm PES}$) for the EMC of (b).}
 \end{figure*}

Next, we deduce the effective charge carrier density $n_{\rm eff}$ from the detailed mapping of the VBs by ARPES.
Since the VB maximum is found 105\,meV below  $E_{\rm F}^{\rm PES}$  (Fig.\ \ref{fig:ARPES_MDC}d), one might conjecture the absence of a Fermi surface, i.e.,  $n_{\rm eff}=0$, at least close to the surface, i.e., at the origin of the ARPES signal.
However, the bands are significantly broadened, such that their tails cut $E_{\rm F}^{\rm PES}$ (Fig.\ \ref{fig:ARPES_MDC}). Hence, the tails of the VB give rise to a non-vanishing $n_{\rm eff}$.
Accordingly, we replace the usual
\begin{equation}
n_{\rm eff} =\frac{2}{8\pi^3}\iiint_{\rm Fermi\,volume}d^3\mathbf{k},
\end{equation}
where the Fermi volume includes all occupied states, with:
\begin{equation}
\label{eq:alpha}
n_{\rm eff} =\frac{2}{8\pi^3}\sum_{j} \iiint_{\rm BZ} \alpha_j(\mathbf{k})\hspace{1mm} d^3\mathbf{k}.
\end{equation}
The integral covers the whole BZ and includes the weight of each state above $E_{\rm F}^{\rm PES}$ (inset of Fig.\ \ref{fig:ARPES_EDC}a) according to
\begin{equation}
\label{eq:alphaj}
\alpha_j(\mathbf{k})=\int_{E_{\rm F}^{\rm PES}}^{\infty}p_{j, \rm norm}(E,\mathbf{k})dE.
\end{equation}
Here, $p_{j, \rm norm}(E,\mathbf{k})$ is the fitted EDC peak at $\mathbf{k}$ of band $j$, after normalizing its area to unity.

Figure \ref{fig:ARPES_EDC}a shows an exemplary EDC (black points) fitted with two Voigt peaks $p_j(E,\mathbf{k})$ (blue and green line), which are multiplied by the Fermi distribution function $f_0(E,T)$ at $T=300$\,K (thin, red line). This provides an excellent fitting result (thick red line). The weights of the two peaks above $E_{\rm F}$ are evaluated to be $\alpha_{\rm 1}(\mathbf{k})= 3$\,\% and $\alpha_{\rm 2}(\mathbf{k})= 1$\,\% (inset of Fig.\ \ref{fig:ARPES_EDC}a) (methods). Generally, we find $\alpha_1(\mathbf{k})\le 7$\,\% for 97\,\% of the EDCs, where the largest $\alpha_1(\mathbf{k})$ are coincident with the maxima of $E_{\rm peak}(\mathbf{k})$. This is illustrated in the inset of Fig.\ \ref{fig:ARPES_EDC}d showing $\alpha_{\rm 1}(\mathbf{k})$ and $\alpha_{\rm 2}(\mathbf{k})$ for the EMC of Fig.\ \ref{fig:ARPES_EDC}b.

We only evaluate the contributions of the two upper VBs ($p_1$, $p_2$), since all other bands are more than 1\,eV below $E_{\rm F}^{\rm PES}$. Then, the $\alpha_j(\mathbf{k})$ for different $\mathbf{k}$ are summed up and multiplied by $1.25$  in order to compensate for the part of the BZ, which is not probed by ARPES. Finally, we normalize appropriately. This eventually leads to \ $n_{\rm eff} =(4.4\pm 1.1)\cdot10^{26}\,\mathrm{m}^{-3}$ with uncertainty resulting from the individual error bars of peak energies and peak widths within the Voigt fits. Taking only the contributions from the upper peak $p_1$, we get $n_{\rm eff,1} =(2.6\pm 0.6)\cdot10^{26}\,\mathrm{m}^{-3}$.  Since the surface might be influenced by band bending, we also calculated $n_{\rm eff}$ for an artificially varying $E_{\rm F}$ with respect to the measured $E_{\rm F}^{\rm PES}$ as displayed in Fig.\ \ref{fig:ARPES_EDC}d.

Next, we compare these $n_{\rm eff}$ with the results from Hall measurements, which yields the bulk charge carrier density $n_{\rm eff, H}= e\sigma_{xy}/B$ ($\sigma_{xy}$: Hall conductivity, $B$: magnetic field) varying between $1.3\cdot 10^{25}\,\mathrm{m}^{-3}$ and $3.0\cdot 10^{26}\,\mathrm{m}^{-3}$ for nominally identical samples (table \ref{tab:MR}, methods). The variation is probably caused by the known, strong sensitivity of GST transport properties to disorder \cite{Siegrist2011}. The temperature dependence of $\sigma_{xy}$ is small within the interval $T=4-300$\,K (changes $<10$\,\%) demonstrating metallic conductivity. The interval of the $n_{\rm eff,H}$ data is marked in Fig.\ \ref{fig:ARPES_EDC}d. The larger $n_{\rm eff,H}$ excellently match $n_{\rm eff}(E_{\rm F}^{\rm PES})$, while the smaller ones are compatible with an $E_{\rm F}$ shifted further upwards. In any case, the tails of the VB provide enough density of states to host the charge carrier density $n_{\rm eff,H}$. We conclude that $E_{\rm F}$ of GST-225 is indeed well above the VB maximum. In turn, we can estimate the required $n_{\rm eff}$ to locate $E_{\rm F}$ at the VB maximum ($E-E_{\rm F}^{\rm PES}=-105$\,meV) to be $n_{\rm eff}\simeq 3\cdot 10^{27}\,\mathrm{m}^{-3}$ ($n_{\rm eff,1}\simeq 2.3\cdot 10^{27}\,\mathrm{m}^{-3}$), i.e., an order of magnitude larger than the highest values found by the Hall measurements. This excludes a significant downwards band bending of the VB towards the surface.

In principle, one could argue that the peak width is not due to disorder, but due to the finite lifetime of the photo-hole produced by ARPES \cite{Huefner,Matsdorf1998}. However, the Voigt fits, which add up a Gaussian peak and a Lorentzian peak, exhibit, on average, 99\,\% (97\,\%) Gaussian contribution and 1\,\% (3\,\%) Lorentzian contribution for $p_1$ ($p_2$). Therefore, the lifetime broadening, encoded in the Lorentzian part, is negligible \cite{Huefner,Matsdorf1998}.
Moreover, the average electron scattering time $\overline{\tau}$ detected by magnetotransport reasonably fits to the disorder induced peak widths (see below).

\subsection*{Electron mean free path from ARPES and magnetotransport}  \label{EMFP}

\begin{table}
\begin{centering}
\begin{tabular}{|c|c|c|c|}
\hline
 $\overline{m}^\star/m_{\rm e}$ & $\overline{k}_0$ $(\mathrm{nm}^{-1})$ & $c$ (nm) & $n_{\rm eff, H}$ $(10^{26}\,\mathrm{m}^{-3})$\tabularnewline
\hline
$0.35\pm 0.01$ & $1.47\pm 0.02$ & 1.04  & $0.13-3.0$\tabularnewline
\hline
\end{tabular}\protect
\caption{VB parameter of epitaxial GST-225: $\overline{m}^\star$:
curvature parameter,  $\overline{k}_0$: average position of the cusp of the M-type VB in ($k_x$, $k_y$) direction, $c$: size of the unit cell of the disordered, cubic Petrov phase along the direction perpendicular to the layers \cite{Nonaka2000}, $n_{\rm eff, H}$: charge carrier density interval according to Hall measurements.}
\label{tab:VBwerte}
\end{centering}
\end{table}
Next, we deduce the average scattering lifetime of the electrons ($\overline{\tau}$) and the average mean free path $\overline{\lambda}_{\rm MFP}$ from the combination of ARPES and magnetotransport.
In Supplementary  Note 2, we show that the longitudinal conductivity $\sigma_{xx}$ and $\sigma_{xy}$ can be straightforwardly related to $\overline{\tau}$ for an isotropic, M-shaped parabolic band in $(k_x,k_y)$ direction with negligible dispersion in $k_z$ direction and without peak broadening. Thus, in line with the ARPES data, we approximate the dispersion as
\begin{equation}
E_{\rm peak}(\mathbf{k})=E_{\rm peak,0}-\frac{\hbar^2}{2m^\star}\cdot \left(k_\parallel-k_0\right)^2
\label{eq:dispersion}
\end{equation}
with $(E_{\rm peak, 0}, k_0)$ being the cusp of the inverted parabola and $m^\star:=\hbar^2(d^2E/dk_\parallel^2)^{-1}$ representing the curvature in radial in-plane direction. This $m^\star$ is different from a universal effective mass of the VB, since the band curvature differs for other $\mathbf{k}$ directions. We obtain (Supplementary Note 2):
\begin{eqnarray}\label{eq:6a}
\sigma_{xx}&=& \frac{ e^2 n_{\rm eff} \overline{\tau}}{2 m^\star}\\ 
\sigma_{xy}&=&en_{\rm eff}/B,
\end{eqnarray}
with $\sigma_{xx}$ being distinct by a factor of $1/2$ from the standard Drude result, which is only valid for an isotropic, parabolic band centered at $\Gamma$. To determine $\overline{\tau}$, we have, hence, to deduce $m^\star$ from ARPES, besides $n_{\rm eff}$. Corresponding parabolic fits to $E_{{\rm peak},1}(\mathbf{k})$, exemplary shown in Fig.\ \ref{fig:ARPES_EDC}b, are executed for all EMCs at different azimuths in ($k_x$, $k_y$) direction and different $k_z$. This leads to the histogram of $m^\star$ values in Fig.\ \ref{fig:ARPES_EDC}c with mean
$\overline{m}^\star=(0.35\pm 0.01)\cdot m_{\rm e}$ ($m_{\rm e}$: bare electron mass, table\ \ref{tab:VBwerte}). 

During the same fit, we naturally get an average $k_0$ as given in table\ \ref{tab:VBwerte} ($\overline{k}_0$) and an average $E_{\rm peak,0}$ being $\overline{E}_{\rm peak,0}-E_{\rm F}^{\rm PES}=-122\pm 3$\,meV. With the determined $\overline{m}^\star$, we can use magnetotransport data and eq.\ \ref{eq:6a} to estimate $\overline{\tau}$. For the sample, where $n_{\rm eff,H}$ fits best to $n_{\rm eff}$ from ARPES (table \ref{tab:MR}, methods), we measured $n_{\rm eff,H} =(3.0\pm 0.2)\cdot 10^{26}\,\mathrm{m}^{-3}$ and $\sigma_{xx}= (6\pm 1)\cdot 10^4\,\mathrm{S}\,\mathrm{m}^{-1}$ (at 300\,K) leading to
$\overline{\tau}=5\pm 1$\,fs (table\ \ref{tab:model}). The variation between different samples grown with the same parameters (methods, table \ref{tab:MR}) is negligible.

Straightforwardly, we can determine other parameters of the dispersion of eq.\ \ref{eq:dispersion} including the Fermi wave vector $\overline{k}_{\rm F}$ and the Fermi velocity $\overline{v}_{\rm F}$, while still neglecting the peak broadening  (Supplementary Note 2):
\begin{eqnarray}
\label{eq:6}
&E_{\rm F}-\overline{E}_{\rm peak,0}=-\frac{\hbar^2} {8\overline{m}^\star}\left( \frac{\pi c n_{\rm eff}}{\overline{k}_0}\right)^2 = -(12\pm 1)\,{\rm\,meV}&\\
&|\overline{k}_{\rm F}-\overline{k}_0|=\sqrt{\frac{2\overline{m}^\star |E_{\rm F}-\overline{E}_{\rm peak,0}|}{\hbar^2}}=(0.34\pm 0.02)\,\mathrm{nm}^{-1}&\\
&\overline{v}_{\rm F}= \frac{\hbar}{\overline{m}^\star}|\overline{k}_{\rm F}-\overline{k}_0| = \frac{\pi \hbar}{2 \overline{m}^\star}\cdot \frac{cn_{\rm eff}}{\overline{k}_0}&\\
&= (1.1\pm 0.1)\cdot 10^5\,\mathrm{m}\,\mathrm{s}^{-1}&\\
&\overline{\lambda}_{\rm MFP}=\overline{v}_{\rm F}\cdot \overline{\tau}=\frac{\pi\hbar c}{e^2\overline{k}_0}\cdot\sigma_{xx} = (0.6\pm 0.1)\,{\rm nm}&
\end{eqnarray}
where $c=1.04$\,nm (table\ \ref{tab:VBwerte}) is the length of the unit cell of the metastable rock salt phase (structure model in Fig.\ \ref{fig:Structure}a$-$c)
perpendicular to the layers. The numerical values are again given for the sample with $n_{\rm eff} =(3.0\pm 0.2)\cdot 10^{26}\,\mathrm{m}^{-3}$ and $\sigma_{xx}= (6\pm 1)\cdot 10^4\,\mathrm{S}\,\mathrm{m}^{-1}$ (methods, table \ref{tab:MR}). $E_{\rm F}$ is located in the band belonging to $p_1$ for all $n_{\rm eff,H}$ of our samples. Note  that neither $\overline{m}^\star$, as usual, nor $n_{\rm eff}$, as typical for two-dimensional (2D) systems \cite{Ando}, enters the evaluation of $\overline{\lambda}_{\rm MFP}$, but only $\overline{k}_0$ does. This reflects the dominating 2D-type dispersion for GST-225.

We also used a more refined, numerical calculation, which considers the variation of $m^\star$ across the BZ and the peak broadening, i.e., the fact that $E_{\rm F}^{\rm PES}$ is above the VB maximum, explicitly. Therefore, we use the low-temperature limit of Boltzmann's relaxation model. We assume that the scattering time $\tau_j(\mathbf{k})$ does not depend on $\mathbf{k}$ and band index $j$, reading $\overline{\tau}:=\tau_j(\mathbf{k})$ which leads to
\begin{eqnarray}
\label{eq:boltz2}
\sigma_{xx} & = \frac{-e^{2}}{4\pi^3}  \iiint  \sum_{j=1}^2 v_{x,j}^{2}(\mathbf{k})  \tau_j(\mathbf{k})p_{j,\rm n}(E,\mathbf{k}) \frac{df_{0}(E (\mathbf{k}),T)}{dE} d^{3}\mathbf{k} \nonumber\\ \\
 & = \frac{\overline{\tau}e^{2}}{4\pi^3}  \iiint  \sum_{j=1}^2 v_{x,j}^{2}(\mathbf{k}) p_{j,\rm n}(E,\mathbf{k}) \delta(E (\mathbf{k})-E_{\rm F}) d^{3}\mathbf{k}\,. \nonumber\\
\end{eqnarray}
The group velocity $\underline{v}_j(\mathbf{k})$ is determined from the ARPES data as $\underline{v}_j(\mathbf{k})=\nabla_{\mathbf{k}}E_{{\rm peak},j}(\mathbf{k})/\hbar$ with the derivative taken at $E_{{\rm peak},j}$ and not at $E_{\rm F}^{\rm PES}$.
Since the results now depend critically on $E_{\rm F}$, we restrict the analysis to the sample with $n_{\rm eff}\simeq n_{\rm eff,H}$ as used in eq.\ \ref{eq:6}.
Numerically, we obtain $\overline{\tau}= 3 \pm 1 $\,fs, which is nearly a factor of two smaller than within the simplified calculation.
By the same numerical $p_{j,\rm n}(E,\mathbf{k})$ weighting,  we determine the average group velocity at $E_{\rm F}^{\rm PES}$ as
$\overline{v}_{\rm F}=(1.2\pm 0.1)\cdot 10^5\,\mathrm{m}\,\mathrm{s}^{-1}$ leading to
$\overline{\lambda}_{\rm MFP}\simeq \overline{v}_{\rm F} \cdot \overline{\tau}=0.4\pm 0.1$\,nm (table\ \ref{tab:model}).

\begin{table}
\begin{centering}
\begin{tabular}{|c||c|c|}
\cline{2-3}
\multicolumn{1}{c||}{} &  Simple model &  Full model\tabularnewline
\hline
\hline
$\overline{\tau}$ in fs & $5\pm 1$ & $3 \pm 1$\tabularnewline
\hline
$\overline{v}_{\rm F}$ in $\mathrm{m}\,\mathrm{s}^{-1}$ & $(1.1\pm 0.1)\times10^{5}$ & $(1.2\pm 0.1)\times10^{5}$\tabularnewline
\hline
$\overline{\lambda}_{\rm MFP}$ in nm & $0.6\pm0.1$ & $0.4\pm0.1$\tabularnewline
\hline
\end{tabular}\protect
\caption{Averaged scattering time $\overline{\tau}$, averaged Fermi velocity $\overline{v}_{\rm F}$ at $E_{\rm F}^{\rm PES}$ and resulting average mean free path $\overline{\lambda}_{\rm MFP}$ for the sample with charge carrier density $n_{\rm eff} =(3.0\pm 0.2)\cdot 10^{26}\,\mathrm{m}^{-3}$ and conductivity $\sigma_{xx}= (6\pm 1)\cdot 10^4\,\mathrm{S}\,\mathrm{m}^{-1}$. The values are deduced within a simple model neglecting the disorder induced peak broadening and the variation of $m^\star$ across the BZ and within the full model taking  both aspects into account.}
\label{tab:model}
\end{centering}
\end{table}
We compare the results of the refined model and the simplified model (eq.\ \ref{eq:6a}, \ref{eq:6}) in table\ \ref{tab:model} revealing that the simplified model returns reasonable values, but deviates from the more exact, refined model by up to 40\,\%. This must be considered for the interpretation of magnetotransport data, where eq.\ \ref{eq:6a} and \ref{eq:6} provide only reasonable estimates for $n_{\rm eff}$, $\overline{v}_{\rm F}$, $\overline{\tau}$, and $\overline{\lambda}_{\rm MFP}$ with an intrinsic error of about 40\,\%.

Finally, we comment on the peak width, which is, on average, $\Delta \overline{E}_1 = 0.20 \pm 0.03$\,eV (FWHM) for the band $p_1$. This can be compared with $\Delta \overline{E}_1 \sim \hbar/\tau$ (Supplementary Note 3). We find $\hbar/\Delta \overline{E}_1  = 3.5$~fs in excellent agreement with
$\overline{\tau}=3$\,fs as deduced from the transport data of the sample with highest conductivity (largest $\overline{\tau}$) (table \ref{tab:model}). This corroborates the assignment of the peak widths to disorder broadening, as already conjectured from its dominating Gaussian shapes. The fact that $\sigma_{xx}$ increases by only 15\,\% between room temperature and $T=4$\,K \cite{Bragaglia2016} additionally shows that $\overline{\tau}$ is dominated by disorder scattering. We conclude that disorder
broadening is responsible for the peak widths within the $E(\mathbf{k})$ spectral function of the upper valence band of GST-225. The relatively large peak widths (0.2\,eV) allows the Fermi level to be well above all peak maxima, i.e., the charge carrier density fits into the tails of the bands. We finally stress that the peak broadening is not the origin of the p-type doping, which has been found previously to be dominated by excess vacancy formation \cite{Edwards2006,Wuttig2007b}.

\subsection*{In-gap surface state} \label{SP}

Motivated by our previous finding, that an M-shaped VB with maxima away from high symmetry points is only compatible with DFT calculations of GST exhibiting nontrivial topology \cite{Pauly2013}, we searched for a surface state within the fundamental band gap. We found such a state by two-photon ARPES (2P-ARPES) exhibiting a linear, largely isotropic dispersion as well as helical, circular dichroism. The state is probably connected to a strongly spin-polarized state at the VB maximum revealed by spin polarized ARPES (S-ARPES).

\begin{figure}
\includegraphics[width=1\linewidth]{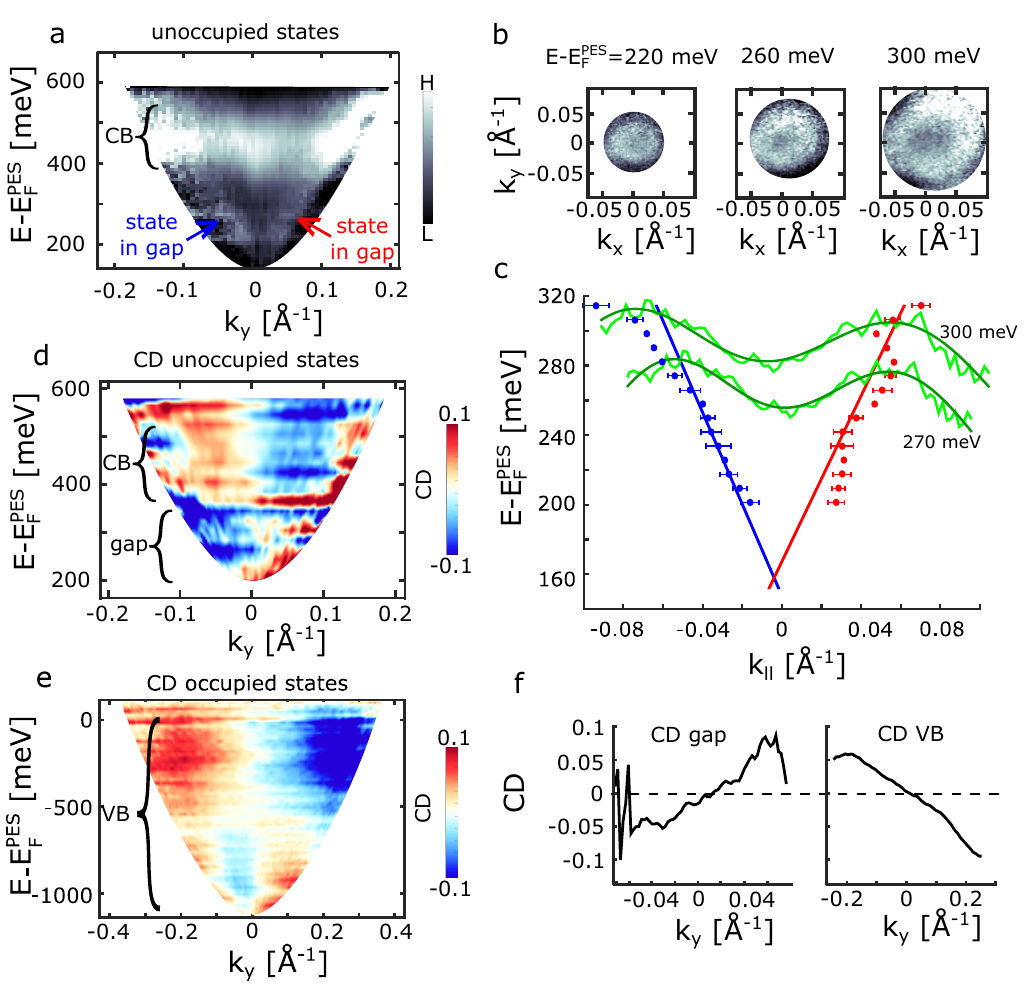}
\caption{\label{fig:2PPE} Two-photon ARPES:
(a) EMC of unoccupied states with energy regions of conduction band (CB) highlighted and presumable topological surface state (state in gap) marked by arrows;
(b) CECs within the bulk band gap for different $E-E_{\rm F}^{\rm PES}$ as indicated;
(c) green lines: MDCs along $k_y$ at $E-E_{\rm F}^{\rm PES}$ as marked (jagged lines) with fits consisting of two Voigt peaks (smooth lines); red, blue dots: peak positions of MDC Voigt fits averaged for the MDCs along $k_x$ and $k_y$; red, blue lines: linear fits to the red and blue dots;
(d) Circular dichroism (CD) intensity of the 2P-ARPES data of (a) with CB and bulk band gap region (gap) marked; 
(e) CD intensity of the VB states recorded by one photon ARPES; 
(f) CD intensity after energy integrating: left: $E-E_{\rm F}^{\rm PES}=200-350$\,meV from (d), right:  $E-E_{\rm F}^{\rm PES}=-580 - 0$\,meV from (e) (VB).}
\end{figure}
Optical measurements revealed a band gap of GST-225 of $E_{\rm gap}\simeq 0.5$\,eV in rough agreement with DFT data \cite{Lee2005}, which was recently corroborated by scanning tunneling spectroscopy ($E_{\mathrm{gap}}=0.45\pm 0.05$\,eV) \cite{Pauly2013,Kellner,Kellner2017}. Hence, we have to probe this energy interval above the VB maximum, which does not contain bulk states.
We employed laser-based 2P-ARPES at pump energy $h\nu=1.63$\,eV and probe energy $h\nu = 4.89$\,eV, hence, populating states in the bulk band gap and in the lower part of the conduction band by the pump, which are subsequently probed by ARPES using the probe pulse. The time delay $\Delta t=1.33$\,ps is chosen to optimize the contrast of the states within the bulk band gap.
The EMC in Fig.\ \ref{fig:2PPE}a reveals a strong band above $E-E_{\rm F}^{\rm PES}=350$\,meV, which we attribute to the bulk conduction band (CB) at 450\,meV above the VB maximum. Below this CB, a mostly linearly dispersing band faintly appears (arrows). Corresponding CECs (Fig.\ \ref{fig:2PPE}b) exhibit a largely circular structure of this band in the $(k_x,k_y)$ plane increasing in diameter with increasing energy. The linear dispersion of the band is deduced by applying two Voigt fits to each MDC as shown for two examples in Figure \ref{fig:2PPE}c. The resulting $E_{\rm peak}(|\overline{\mathbf{k}}|)$ (points), averaged along $k_x$ and $k_y$, are fitted using $E_{\rm peak}(|\overline{\mathbf{k}}|)=E_{\rm peak, D}\pm \hbar v_{\rm D} k_\|$ (red and blue line). This reveals a presumable band crossing at $E_{\mathrm{peak, D}}-E_{\rm F}^{\rm PES}= 160\pm10$\,meV and a band velocity $v_{\mathrm{D}}=(3.8\pm 0.3)\times 10^{5}\,\mathrm{m}\,\mathrm{s}^{-1}$.

Due to the relatively strong one-photon background (methods), we could not evaluate the 2P-ARPES signal at lower $E-E_{\rm F}^{\rm PES}$, such that the presumable crossing point was not probed directly. However, all  signatures of this band are compatible with a TSS with mostly linear dispersion.

In addition, we probed the circular dichroism (CD) by 2P-ARPES using a linearly polarized pump and a circular polarized probe pulse \cite{Niesner2012}. The CD intensity is the scaled difference of photoelectron intensity after clockwise and counterclockwise circular polarization of the probe. It is known that the CD cannot be directly assigned to a spin polarization of initial states,\cite{Scholz2013} but is likely related to an interplay of spin and orbital textures \cite{Crepaldi2014}.
In our case, it shows a sign inversion with the sign of $\mathbf{k}$ (Fig.\ \ref{fig:2PPE}d). The opposite inversion is found within the CB and the upper VB, the latter probed by CD measurements of conventional ARPES (Fig.\ \ref{fig:2PPE}e). The same sequence of CD inversions between VB, TSS, and CB has been found for the prototype strong topological insulators Bi$_2$Se$_3$\cite{Wang2013}, Bi$_2$Te$_3$ \cite{Scholz2013,Wang2013}, and Sb$_2$Te$_3$ \cite{Seibel2015}, which
is an additional hint that the linearly dispersing state within the bulk band gap is a TSS.

\begin{figure}
\includegraphics[width=1\linewidth]{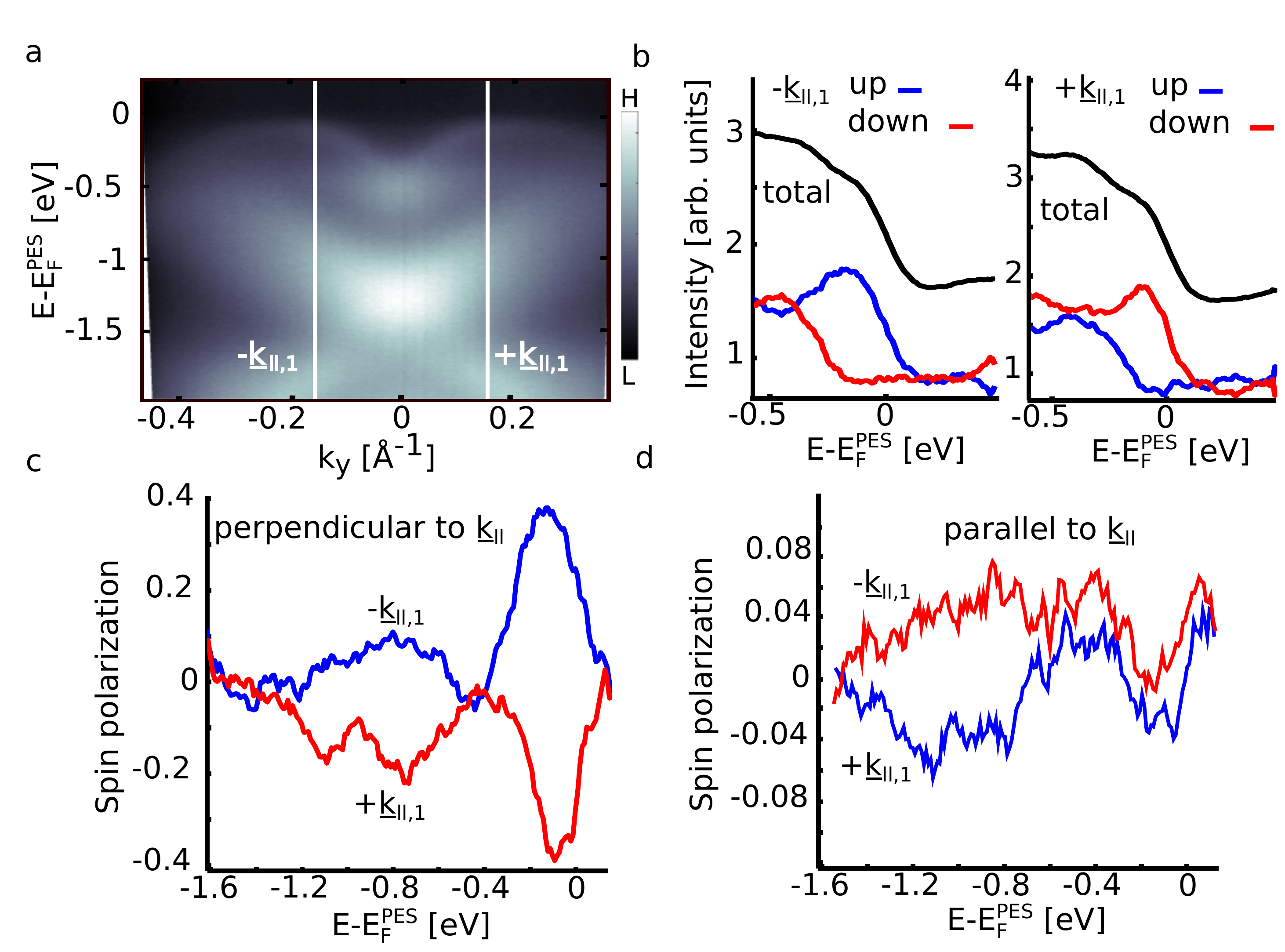}
\caption{\label{fig:ARPES_SPIN} Spin polarized ARPES: (a) ARPES spectrum along  $k_y$ ($\Gamma$K) at $h\nu=30$\,eV; labeled white lines mark the $k_y$ positions for the spin resolved measurements in (b)$-$(d); (b) recorded in-plane signal of the spin up (blue curve) and spin down (red curve) channel perpendicular to $\mathbf{k}_{\parallel,1}$ after scaling with the Sherman function; black line: sum of red and blue line; left: $-\mathbf{k}_{\parallel,1}$, right:  $+\mathbf{k}_{\parallel,1}$; (c) resulting in-plane spin polarization perpendicular to  $\mathbf{k}_{\parallel,1}$; (d) in-plane spin polarization parallel to $\mathbf{k}_{\parallel,1}$. Note the different scales in (c) and (d).}
\end{figure}
Another fingerprint of non-trivial surface states is spin polarization \cite{Hasan2008,Qi2011,Ando2013}.
Such spin polarized surface states have been predicted by DFT calculations of the cubic Petrov phase of GST-225, in particular, a TSS traversing the band gap and a Rashba-type surface state at $E-E_{\rm F}^{\rm PES}=-300$ ... $-700$\,meV \cite{Pauly2013}. To this end, we probe the spin polarization of the occupied states at a selected pair of in-plane wave vectors $\pm \mathbf{k}_{\parallel,1}$ (Fig.\ \ref{fig:ARPES_SPIN}). We choose $h\nu=30$\,eV such that the CEC of the bulk VB is large in diameter (Fig.\ \ref{fig:ARPES_MDC}f), thereby increasing the possibility to probe a surface state in the inner part of the BZ, where DFT of the cubic Petrov phase predicted the presence of a TSS \cite{Pauly2013}. Moreover, we used $|\mathbf{k}_{\parallel,1}|=0.17 \pm 0.04$\,\AA$^{-1}$, large enough to avoid overlap of intensity from $\mathbf{k}_{\parallel,1}$ and $-\mathbf{k}_{\parallel,1}$, thereby getting along with the typically reduced angular resolution of S-ARPES. Indeed, we find strong in-plane spin-polarization of 40\,\% close to $E_{\rm F}^{\rm PES}$ (Fig.\ \ref{fig:ARPES_SPIN}b,c). The spin polarization inverts sign with the sign of the in-plane wave vector and is perpendicular to $\mathbf{k}_{\parallel,1}$ within error bars (Fig.\ \ref{fig:ARPES_SPIN}c,d). The other spin polarized state at lower $E-E_{\rm F}^{\rm PES}=-0.5$ ... $-1.2$\,eV might be related to the Rashba state mentioned above, which has similarly been found, e.g., for Sb$_2$Te$_3$(0001) \cite{Pauly2012}.

The peak energy of the spin polarized state close to $E_{\rm F}^{\rm PES}$ is $E_{\rm peak}-E_{\rm F}^{\rm PES}=-120\pm 40$\,meV (Fig.\ \ref{fig:ARPES_SPIN}b), i.e., very close to the VB maximum, such that likely the spin-polarized state extends into the band gap. The error mostly comes from the different peak energies at $+\mathbf{k}_{\parallel,1}$ and $-\mathbf{k}_{\parallel,1}$. We cannot prove that this state is connected to the linearly dispersing state of Fig.\ \ref{fig:2PPE}, which would hit the VB maximum at $|\mathbf{k}_{\parallel}|=0.10$\,\AA$^{-1}$, if perfectly linear in dispersion down to the VB , but we believe that this is likely.\\
One might ask  why such a linearly dispersing state is not observed in the one-photon ARPES data. A possible explanation is the fact that a surface state will follow the roughness of the surface, which for our films amounts to angles of 0.5$^\circ-$~3$^\circ$ according to atomic force microscopy \cite{Pauly2013}.
Assuming, for the sake of simplicity, the same state dispersion on all surfaces, this results in a $k_\| $ broadening of the surface state by $(0.15-1.5)\,\mathrm{nm}^{-1}$ for $E_{\rm kin}=12-27$\,eV. The widths of the 2P-ARPES peaks in Fig.\ \ref{fig:2PPE}c is $0.6\,\mathrm{nm}^{-1}$ and, hence, well compatible with this analysis. Since the bulk states are not influenced by this broadening mechanism, it gets rather difficult to discriminate the TSS in the presence of bulk VB states at similar $E(\mathbf{k})$ as within our one-photon ARPES data.

We did not reproduce the dispersion of the found state in the bulk band gap by the DFT calculations of slabs of the cubic Petrov phase, which revealed a less steep dispersion of its TSS and another $E_{\rm D}$ \cite{Pauly2013}. We ascribe this discrepancy to the known, strong sensitivity of the TSS to details of GST's atomic structure \cite{Kim2010,Sa2011,Kim2012,Sa2012,Silkin2013,Pauly2013}. However, besides these remaining questions, both, the strong spin polarization close to the VB maximum and the linear dispersion within the bulk band gap are compatible with a TSS. This corrobarates the previous conjecture of a topologically inverted band structure of metastable GST-225 \cite{Pauly2013}.

\subsection*{Possible contribution of the topological surface state to conductivity} \label{TSS}

The possible presence of a TSS at $E_{\rm F}^{\rm PES}$, naturally protected from backscattering \cite{Hasan2008,Qi2011,Ando2013}, raises the question whether it would contribute significantly to the conductivity. To answer this, we firstly compare the charge carrier density of the presumable TSS $ n_{\rm 2D,TSS}$ with the measured charge carrier density of the epitaxial film, after projecting to 2D according to $n_{\rm 2D,H}=n_{\rm eff,H}\cdot d$ ($d$: film thickness). The latter varies between $n_{\rm 2D,H} = 7\cdot 10^{18}\,\mathrm{m}^{-2}$ (film of largest conductivity) and $n_{\rm 2D,H} = 3\cdot 10^{17}\,\mathrm{m}^{-2}$ (table \ref{tab:MR}, methods). For the non-degenerate 2D band of a linearly dispersing TSS, we have\cite{Ando}:
\begin{equation} \label{eq:4}
 n_{\rm 2D,TSS}= \frac{k^{2}_{\rm F}}{2\pi}.
\end{equation}
A reasonable assumption for $k_{\rm F}$ results from extrapolating the fitted linear dispersion of Fig.\ \ref{fig:2PPE}(c) to $E_{\rm F}^{\rm PES}$ leading to $|k_{\rm F}|=6\cdot 10^8\,\mathrm{m}^{-1}$. An upper estimate is $|k_{\rm F}|=1.7\cdot10^9\,\mathrm{m}^{-1}$, i.e., the $k_\parallel$ value of the spin polarized state at the VB maximum. Hence, we get $n_{\rm 2D,TSS} \simeq 6\cdot 10^{16}\,\mathrm{m}^{-2}$, respectively $n_{\rm 2D,TSS} \le 5\cdot 10^{17}\,\mathrm{m}^{-2}$.

Comparing with the sample exhibiting $n_{\rm eff}\simeq n_{\rm eff, H}$ ($n_{\rm 2D,H} = 7\cdot 10^{18}\,\mathrm{m}^{-2}$), $n_{\rm 2D,H}$ is more than an order of magnitude larger than $n_{\rm 2D,TSS}$. We conclude that the charge carrier density is dominated by the bulk VB.

However, the mobility of a TSS ($\mu_{\rm TSS}$) could be much larger than the mobility of the bulk VB ($\mu_{\rm bulk}$). Such a TSS conductivity dominates, if $\mu_{\rm TSS}/\mu_{\rm bulk} > n_{\rm 2D, bulk}/n_{\rm 2D,TSS}$ ($n_{\rm 2D,bulk}$: charge carrier density in the bulk VB after 2D projection).
Within the two band model, we have \cite{Ando}
\begin{eqnarray}
\label{eq:2band}
n_{\rm 2D,H}&=&\frac{n_{\rm 2D, TSS}^2\cdot \mu_{\rm TSS}+n_{\rm 2D, bulk}^2\cdot \mu_{\rm bulk}}{n_{\rm 2D, TSS}\cdot \mu_{\rm TSS}+n_{\rm 2D, bulk}\cdot \mu_{\rm bulk}}\\
\sigma_{xx}&=&(n_{\rm 2D, TSS}\cdot \mu_{\rm TSS}+n_{\rm 2D, bulk}\cdot \mu_{\rm bulk})\cdot \frac{e}{d}\,.
\end{eqnarray}
We evaluate these equations for the sample with $n_{\rm eff}\simeq n_{\rm eff, H}$ ($\sigma_{xx}=6\cdot10^4\,\mathrm{S}\,\mathrm{m}^{-1}$) using the assumption of a linearly dispersing TSS down to $E_{\rm F}^{\rm PES}$ ($n_{\rm 2D,TSS} = 6\cdot 10^{16}\,\mathrm{m}^{-2}$). We solve eq.\ \ref{eq:2band} and  $n_{\rm 2D, TSS}\cdot \mu_{\rm TSS} \ge n_{\rm 2D, bulk}\cdot \mu_{\rm bulk}$ for the three remaining unknowns ($\mu_{\rm TSS}$, $\mu_{\rm bulk}$, $n_{\rm 2D, bulk}$) leading to $\mu_{\rm TSS}\ge 0.1\,\mathrm{m}^2 (\mathrm{Vs})^{-1}$. The threshold for dominating $\mu_{\rm TSS}$ is even lower for the other samples (table \ref{tab:MR}, methods). We moreover assume that only the surface contains a highly mobile TSS. In turn, the threshold for dominating $\mu_{\rm TSS}$ has to be divided by two, if the interface to the Si(111) contains a TSS with the same $n_{\rm 2D,TSS}$ and $\mu_{\rm TSS}$. For comparison, the record mobilities found for TSS in other systems (Bi$_2$Se$_3$, BiSbTeSe$_2$) are $\mu_{\rm TSS}\simeq 1\,\mathrm{m}^2 (\mathrm{Vs})^{-1}$\cite{Oh2015,Xu2016}, i.e., significantly larger than the threshold.

We conclude with the encouraging possibility to prepare highly mobile, metastable GST-225 films, noting that polycrystalline films exhibit  $\sigma_{xx}<6400\,\mathrm{S}\,\mathrm{m}^{-1}$ \cite{Volker}, which is an order of magnitude lower than for our best epitaxial film. Thus, one might boost the GST-225 conductivity by the combination of epitaxial films and adequate interface design leading to optimized TSS mobility \cite{Oh2015,Xu2016}. This could be exploited within innovative devices combining the fast \cite{Yamada1991,Loke2012} and energy efficient \cite{Xiong2011} phase change with ultrahigh mobility of the on-state.

\section*{Summary}

We have mapped the 3D electronic bulk band structure $E(\mathbf{k})$ close to $E_{\rm F}$ of epitaxial GST films in the metastable rock salt phase and have correlated the results with magnetotransport data of identically prepared samples. The constant energy surfaces of the valence band close to $E_{\rm F}$ are hexagonal tubes with little dispersion along $k_z$, the direction perpendicular to the chemically distinct layers. The valence band maximum is about 100\,meV below $E_{\rm F}$, such that only the tails of the disorder broadened $E(\mathbf{k})$ states contribute to the conductivity. This is in line with the measured charge carrier densities from Hall measurements. We use the mapped band structure in combination with magnetotransport to determine the elastic scattering time (3\,fs) and the mean free path (0.4\,nm), the former being compatible with the peak widths found in ARPES.

Our detailed modeling reveals that variations of the band structure across the BZ. i.e., different band curvatures and peak broadenings, modify the deduced scattering time and average mean free path by about 40\,\%, such that simplified models, as typically used for the interpretation of magnetotransport data, cannot provide a better accuracy.

Besides, we find a linearly dispersing state within the bulk band gap which might have a topological origin. We estimate that this state would dominate the longitudinal conductivity at a mobility above $0.1\,\mathrm{m}^2 (\mathrm{Vs})^{-1}$, which is lower than the best mobilities of topological surface states so far ($\mu \simeq 1\,\mathrm{m}^2 (\mathrm{Vs})^{-1}$) \cite{Oh2015,Xu2016}. Currently, topological conductivity is not expected to be dominant, but by surface or interface optimization one might exploit it in future GST devices providing an ultrahigh mobility on-state.

\section*{Methods}

{\bf Sample preparation.} The GST films are grown by MBE at base presure  $10^{-8}$\,Pa on a Si(111) substrate using elementary sources of Ge, Sb and Te and a substrate temperature of $250\,^{\circ}$C. The growth rate was $0.05\pm0.02\,\mathrm{nm}\,\mathrm{s}^{-1}$ and the pressure increased to $2\cdot 10^{-7}$\,Pa during growth. XRD reveals that the films grow epitaxially in the single crystalline, metastable rock salt phase with [111] surface. The surface is Te terminated as evidenced by DFT calculations (not shown). The film thickness is determined by XRD fringes or by X-ray reflectometry to be 25\,nm, 18\,nm, and 13\,nm for the samples used for ARPES, 2P-ARPES and S-ARPES, respectively.
Twin domains are found, i.e., adjacent areas of ABC and CBA stacking of the hexagonal layers \cite{Pauly2013}. A peak indicating the formation of a vacancy layer is observed by XRD, hinting to more ordered samples than in the purely disordered rock salt phase \cite{Bragaglia2016}.
The XRD data recorded after the ARPES measurements show variations in the (222) peak position by up to 1.5\,\% \cite{Bragaglia2016} and in the height of the vacancy layer peak by 15-25\,\%. However, the ARPES data of these samples are quite similar, i.e., peak positions of the VB vary by less than the peak widths.
Samples are transferred in UHV between the MBE and the three different, analyzing ARPES systems using a UHV shuttle with background pressure of $p = 5\times10^{-10}$\,mbar. This prevents oxidation and surface contamination as cross-checked by x-ray photoelectron spectroscopy (XPS), such that no further preparation steps are required. The UHV transfer is crucial, since surface oxidation starts already at $\sim 1$\,Pa$\cdot$s of O$_2$ \cite{Yashina2008}.

{\bf Photoelectron spectroscopy.} The ARPES measurements of the valence band are recorded at a sample temperature $T=300$\,K at BESSY II (beam line UE112-lowE-PGM2 ($1^2$)) using a Scienta R8000 analyzer with energy resolution 20\,meV and angular resolution $0.2-0.3^\circ$. Linearly p-polarized light with photon energies $h\nu=16-31$\,eV and an incidence angle of $\sim 45^\circ$ is applied, which enabled a three-dimensional mapping of the band structure in momentum space ($k_{x}$, $k_{y}$, $k_{z}$). The Fermi energy $E_{\rm F}^{\rm PES}$ of the ARPES setup has been determined on polycrystalline Cu with 5\,meV precision.

The data set contained $998\times666\times49\times29$ pixels, i.e., 998 different photoelectron energies $E_{\rm kin}$, 666 different azimuthal angles $\phi$, 49 polar angles $\theta$, and 29 photon energies $h\nu$. The energy interval $E_{\rm kin} \in [E_{\rm F}+0.042$\,eV,~$E_{\rm F}+0.104$\,eV] is used for background subtraction for each energy distribution curve (EDC) at a particular ($\phi$, $\theta$, $h\nu$). Subsequently, the data are smoothed along $E_{\rm kin}$ and $\phi$ by a 5-point averaging. Accordingly, the data set is reduced to $256\times256\times49\times29$ pixels. Finally, all EDCs are scaled to the same average value for each ($\phi$, $\theta$, $h\nu$).

In order to deduce band centers $E_{{\rm peak},j}(\mathbf{k})$ of band $j$, MDCs and EDCs at constant $h\nu$ are extracted from the data and fitted by two or four Voigt peaks with variable intensities, widths, and relative contributions of the Gaussian and the Lorentzian. This leads to an excellent fit quality with negligible residuals as exemplarily shown in Fig.\ \ref{fig:ARPES_MDC}c, Fig.\ \ref{fig:ARPES_DFT_MDC}g-h, and Fig.\ \ref{fig:ARPES_EDC}a. The resulting up to four $(k_x,k_y)_i$ of MDC fits are then attributed to the preselected $E-E_{\rm F}^{\rm PES}=E_{\rm peak}$, respectively, the resulting $E_{{\rm peak},j}$ of EDC fits are attributed to the preselected $(k_x,k_y)$ values. The resulting $E_{{\rm peak}, j}(k_x,k_y)$ curves deduced from the two methods vary by $\Delta k_\parallel \simeq 0.01$\,\AA$^{-1}$, respectively, by $|\Delta E_{\rm peak}|=5-15$\,meV, except for extreme $\mathbf{k}$ values (see main text). The small deviations contribute straightforwardly to the error of the determined effective charge carrier densities $n_{\rm eff}$ and curvature parameters $m^\star$ (Fig.\ \ref{fig:ARPES_EDC}).

Displaying the upper $E_{\rm peak,1}(k_x,k_y)$ at selected $E_{\rm peak}$ for different $h\nu$, as shown for energy $E_{\rm peak}=E_{\rm F}^{\rm PES}$ in Fig.\ \ref{fig:ARPES_MDC}f, consistently reveals a minimum diameter of the resulting constant energy lines at $h\nu\simeq 21$\,eV. Since DFT finds the minimum diameter of the upper VBs at the BZ boundary (e.g., Fig.\ \ref{fig:ARPES_MDC}h), we assume that the minimum at $h\nu\simeq 21$\,eV corresponds to the BZ boundary in $k_z$ direction.
This assumption is used to determine the inner potential $E_{\rm inner}$ with respect to the vacuum level for the final state electrons in the crystal
according to $E_{\rm final}=\hbar^2 |\mathbf{k}|^2/2m_{\rm e}-E_{\rm inner}$.
Restricting $E_{\rm inner}$ between 10\,eV and 25\,eV leaves us with the only possibility of $E_{\rm inner}=14$\,eV. However, if the minimum diameter is in the center of the BZ, we would get $E_{\rm inner}=20$\,eV. Since these differences are not important for our main conclusions, we select the most reasonable assumption that the smallest diameter is at the BZ boundary.
Using the inner potential, we calculate $k_z$ according to $k_z = 1/\hbar \cdot \sqrt{2m_e E_{\rm kin} \cos{(\theta)}^2+E_{\rm inner}}$. For the CECs and CESs in Fig.\ \ref{fig:ARPES_MDC}, we use an average value of $\theta$ to relate $h\nu$ to $k_z$.\\
The ARPES data cover only 80\,\% of the BZ , i.e., a small part in $k_{z}$ direction is missing (Fig.\ \ref{fig:ARPES_MDC}e). This is due to the fact, that at lower and higher $h\nu$, the ARPES intensity drops drastically, such that fits become unreliable. However, in line with the DFT results, we do not believe that the hexagons change strongly within the remaining 20\,\%.

{\bf Fit procedures and fit errors} 
All peaks of MDC and EDC curves are fitted by several Voigt peaks, i.e., by a combination of a Gaussian and a Lorentzian peak with the same maximum each. Comparing the results of MDC fits and EDC fits for energies below the VB maximum reveals only small differences between deduced $E(\mathbf{k})$ values by $0.1\,\mathrm{nm}^{-1}$ or 10\,meV on average, except for the extreme cases $k_\parallel<0.5\,\mathrm{nm}^{-1}$ and $k_\parallel>2.5\,\mathrm{nm}^{-1}$ ($k_\parallel:=\sqrt{k_x^2+k_y^2}$). The small discrepancies set a lower bound for error margins.

In order to extract $n_{\rm eff}$ from the fitted peaks of EDCs, the peak areas of $p_j(E, \mathbf{k})$ ($j=1,2$) are normalized to one leading to $p_{j,\rm norm}(E,\mathbf{k})$, from which we evaluate the relative part of the peaks above $E_{\rm F}^{\rm PES}$ (inset of Fig.\ \ref{fig:ARPES_EDC}a), being $\alpha_j(\mathbf{k})$, the unoccupied percentage of the corresponding $E(\mathbf{k})$ state.

The error of $\overline{m}^\star$ is only slightly smaller than the error of individual $m^\star$, being 5\,\% on average, which is due to the considerable variation of $m^\star$ across the BZ (Fig.\ \ref{fig:ARPES_EDC}c). The deviation of individual curves from the parabola is negligible (Fig.\ \ref{fig:ARPES_EDC}b), i.e., the average energy distance of individual $E_{\rm peak, 1}(\mathbf{k})$  from the parabola ($\approx 8$\,meV) is less than the average fit error from the determination of $E_{\rm peak,1}(\mathbf{k})$ by Voigt fits ($\approx 30$\,meV).

{\bf Spin polarized photoelectron spectroscopy.} Spin resolved ARPES measurements are conducted at BESSY II, too, using the electron analyzer SPECS PHOIBOS 150 and linearly p-polarized synchrotron radiation at $h\nu=30$\,eV and incidence angle $45^\circ$ at $T=300$\,K, providing an energy resolution of 100\,meV and an angular resolution of $1-2^\circ$. Spin analysis is performed with a Rice University Mott polarimeter operated at 26\,kV resulting in a Sherman function of $S=0.118$.

{\bf Two photon photoelectron spectroscopy.} Angle-resolved bichromatic 2P-ARPES and additional, conventional ARPES experiments are conducted using the first, third and fourth harmonic of a titanium:sapphire oscillator, i.e., $h\nu=1.63$\,eV, $h\nu=4.89$\,eV, and $h\nu=6.2$\,eV, within a home-built setup \cite{Thomann1999,Niesner2012}. The repetition rate of the laser is 80\,MHz and the pulse length is 166\,fs.
The beam is initially p-polarized with an incidence angle of $60^{\circ}$. The photon energy $h\nu=1.63$\,eV is used for the pump pulse followed by the probe pulse at $h\nu=4.89$\,eV, which hits the sample at a time delay $\Delta t$ after the pump. Due to time restrictions, $h\nu$ of the probe pulse has not been changed such that we probe only a single $k_z$. The photon energy $h\nu=6.2$\,eV is used for conventional ARPES to cross check the results obtained at BESSY II. Circular polarization, necessary for circular dichroism (CD) experiments, is obtained using a $\lambda/4$ wave plate. Two-dimensional momentum distribution patterns at constant $E_{\rm kin}$ are recorded using an ellipsoidal 'display-type' analyzer exhibiting an energy and angular resolution of $55~$meV and $3^{\circ}$, respectively \cite{Rieger1983,Niesner2012}. The work function of GST-225 turned out to be $5.1\pm0.1$\,eV leading to a strong one-photon photoemission background from the probe pulse. In order to discriminate 2P-ARPES data from this background, the intensity of measurements at $\Delta t =-2.67$\,ps is subtracted from the data recorded at $\Delta t = +1.33$\,ps. Subsequently, the data are normalized to compensate for inhomogeneities of the channel plates. CD intensity displays the difference of photoelectron intensity using clock wise and counterclockwise polarized probe pulses divided by the sum of the two intensities.

{\bf Magnetotransport.} Magnetotransport measurements are performed \textit{ex-situ}.  Since Hall measurements require insulating substrates while ARPES requires a conducting sample, the Hall data are from samples with lower substrate doping, but grown with identical parameters. After growth they are capped by Te to protect from oxidation. The samples are cut in square shapes of $5\times5\,\mathrm{mm}^{2}$ and, after decapping by a HF dip, are contacted by In and Au bond wires in a four-contact van der Pauw geometry. Magnetotransport measurements are conducted at $T=4-300$\,K, current $I=10$\,mA, and magnetic field $B=0.25$\,T perpendicular to the surface. This leads to  charge carrier densities $n_{\rm eff, H}$ and longitudinal conductivities $\sigma_{xx}$ as displayed in table~\ref{tab:MR} for $T=300$\,K.
It is likely that the upper 5 nm of the sample are oxidized
\cite{Yashina2008,Zhang2010,Gourvest2012}
resulting in a systematic error of 25 \% . We find a relatively broad statistical distribution of  $n_{\rm eff, H}$ and $\sigma_{xx}$, but due to correlations between the two values, the variation of the mobility $\mu=\sigma_{xx}/n_{\rm eff, H}e$ is relatively small.
\begin{table}
\begin{centering}
\begin{tabular}{|c|c|c|c|}
\hline
$n_{\rm eff, H}$& $\sigma_{xx}$ & $\mu$ & $d$ \tabularnewline
$\left( 10^{26}\,\mathrm{m}^{-3}\right)$& $\left( \mathrm{S}\,\mathrm{m}^{-1}\right)$ & $\left( \mathrm{m}^2\,(\mathrm{Vs})^{-1}\right)$ & (nm) \tabularnewline
\hline
\hline
0.13 & 2.700 & 0.0013 & 19 \tabularnewline
\hline
0.26 & 4.000 & 0.0010 & 28 \tabularnewline
\hline
3.0 & 60.000 & 0.0012 & 23 \tabularnewline
\hline
\end{tabular}\protect
\caption{Charge carrier density $n_{\rm eff, H}$, longitudinal conductivity $\sigma_{xx}$, and mobility $\mu$, as determined by 4-point magnetoconductance measurements at $T=300$\,K, as well as thickness $d$ for different epitaxial GST-225 films.}
\label{tab:MR}
\end{centering}
\end{table}

{\bf Band structure calculations.} Density functional theory (DFT) calculations are performed within the generalized gradient approximation. We employ the full-potential linearized augmented plane-wave method in bulk and thin-film geometry as implemented in the FLEUR code. According to Ref. \cite{Sun2006}, the Petrov stacking sequence (Te-Sb-Te-Ge-Te-Vc-Te-Ge-Te-Sb-) \cite{Petrov1968} is assumed for the metastable rock salt phase by tripling the Petrov-type unit cell containing 10 layers in order to realize the ABC stacking of the rock salt phase (Fig.\ \ref{fig:Structure}d). The resulting BZ of the unit cell of 30 layers, is a factor of five smaller in $k_{z}$ direction ($\Delta k_z =0.12$\,\AA$^{-1}$) than the BZ of the disordered metastable rock salt phase, relevant for the ARPES data ($6$ layers in a unit cell, Fig.\ \ref{fig:Structure}c). Hence, we use fivefold backfolding of the experimental data (Fig.\ \ref{fig:ARPES_MDC}g) to compare with the DFT data (Fig.\ \ref{fig:ARPES_MDC}h).
Additional DFT calculations are performed for disordered slabs (Fig.\ \ref{fig:Structure}b) with methodology otherwise similar to Ref. \cite{Deringer}. We simulate maximum disorder by occupying each cationic plane randomly with Ge:Sb:Vc in a 2:2:1 ratio.  These planes are parallel to the (111) surface, and include the disordered subsurface layer, whereas the surface itself is terminated by Te \cite{Deringer}. Different structure models of the cationic plane were randomly generated, and after relaxation showed a standard deviation of 3\,meV/atom in total energies. The computed surface energies range from 12 to 17\,meV\,\AA$^{-2}$ in a Te-poor environment, which can well be reconciled with previous results for ideally ordered GST \cite{Deringer}.\\
More details including atomic coordinates are given in Supplementary Note 4.

\begin{acknowledgments}
We gratefully acknowledge helpful discussions with H.\ Bluhm, M.\ Wuttig, and C.\ M.\ Schneider as well as financial support by the German Science Foundation (DFG): SFB 917 via Project A3, SPP 1666 (Topological Insulators) via Mo858/13-1, and Helmholtz-Zentrum Berlin (HZB). V.L.D.\ was supported by the German Academic Scholarship Foundation. Computing time was provided to G.B.\ by the J{\"u}lich-Aachen Research Alliance (JARA-HPC) on the supercomputer JURECA at Forschungszentrum J{\"u}lich.
\end{acknowledgments}

\section*{Author contributions}
M.M. provided the idea of the experiment. J.K., M.L.\ and C.P.\ carried out all (S)ARPES experiments under the supervision of E.G., J.S-B., O.R., and M.M.. S.O., J.K., P.K., and P.B.\ performed the 2P-ARPES measurements under the supervision of T.F..
J.E.B., R.N.W., S.C., and V.B.\ grew the samples via MBE, supervised by R.C.. V.B.\ performed the electrical transport measurements also supervised by R.C..
G.B.\ provided the DFT calculations for comparison to the ARPES data. V.L.D.\ and R.D.\ performed the additional DFT modelling of the disordered, metastable surfaces. M.L., J.K, and S.O.\ evaluated the experimental data. M.L., M.M., and J.K.\ derived the models for the interpretation of the data as discussed in the manuscript. M.M., M.L.\ and J.K.\ wrote the manuscript containing contributions from all co-authors. The authors declare no competing financial interests.

\section*{Supplementary Note 1: Comparison between ARPES data and DFT data of the cubic Petrov phase}
\begin{table*}
\begin{centering}
\begin{tabular}{|c|c|c|c|c|c|c|}
\hline
$E-E_{\rm F}^{\rm PES}$ & \multicolumn{2}{|c|}{$-200$~meV} &  \multicolumn{2}{|c|}{$-250$~meV} &  \multicolumn{2}{|c|}{$-300$~meV}  \tabularnewline
\hline
\hline
Direction & $k_x$ & $k_y$ & $k_x$ & $k_y$ & $k_x$ & $k_y$\tabularnewline
\hline
ARPES I & $0.22\pm0.05$ & $0.20\pm0.02$ & $0.25\pm0.05$ & $0.22\pm0.04$ & $0.28\pm0.09$ & $0.25\pm0.03$\tabularnewline
\hline
DFT 1 & $0.27$ & $0.21$ & $0.30$ & $0.24$ & $0.33$ & $0.26$\tabularnewline
\hline
DFT 2 & $0.19$ & $0.18$ & $0.24$ & $0.19$ & $0.27$ & $0.24$\tabularnewline
\hline
DFT 3 &  &  & $0.17$ & $0.16$ & $0.20$ & $0.18$\tabularnewline
\hline
ARPES II & $0.08\pm0.06$ & $0.08\pm0.06$ & $0.07\pm0.05$ & $0.08\pm0.07$ & $0.06\pm0.08$ & $0.07\pm0.09$\tabularnewline
\hline
DFT 4 &  &  & & $0.13$  & $0.12$ & $0.10$\tabularnewline
\hline
DFT 5 & $0.19$ & $0.18$ & $0.12$ & $0.12$ & $0.09$ & $0.09$\tabularnewline
\hline
DFT 6 & $0.13$ & $0.13$ & $0.10$ & $0.10$ & $0.08$ & $0.07$\tabularnewline
\hline
\end{tabular}\protect
\caption{Deduced $|k_x|$ and $|k_y|$ values of the hexagonal tubes $E_{\rm peak}(\mathbf{k})$ of Fig.\ 3 of the main text given in \AA$^{-1}$. The labeling of the bands is marked in Fig.\ 3c, f of the main text. The $\pm$ intervals for the ARPES data capture the full dispersion of the corresponding band along $k_z$ (at the given $k_x$ or $k_y$). Notice that DFT 2 and DFT 5 are identical for $E-E_{\rm F}^{\rm PES}=-200$\,meV, since these bands form closed CESs between the tubes DFT 1 and DFT 6, where we give the center of these CESs only.
\label{tab:comparison}}
\par\end{centering}
\end{table*}
As visible in Fig.\ 3 of the main text, there are differences between DFT and ARPES results.  Besides the mostly excellent fit quality of the ARPES data (Fig.\ 3g-h), there are more distinct bands in the DFT calculation (red, blue and green contours), which we tentatively attribute to the higher degree of order in the cubic Petrov phase with respect to the more disordered rock-salt phase probed by ARPES. The different bands in DFT can be attributed mostly to Sb p-states with different nodal structure along the large unit cell of the Petrov phase in $k_z$ direction. We assume that these different bands are differently sensitive to the arrangement of the pure Sb layers, which are only present in the idealized Petrov phase, but not in the experiment. The resulting stronger dispersion along $k_z$ (of a hypothetical rock-salt BZ) even induces closed CESs between the tubes at higher $E-E_{\rm F}^{\rm PES}$ (Fig.\ 3d), which are not observed in the experiment (Fig.\ 3a). Additionally, the inner constant energy surfaces (CESs) of the ARPES data and the DFT data are different.
Corresponding $|\mathbf{k}|$ values for the up to 6 different CESs from DFT and the 2 CESs from ARPES are shown in Supplementary Table \ref{tab:comparison}. While the outer, experimental CES (ARPES I) reasonably fits with the outer DFT CESs (DFT $1-3$), the inner, experimental CES (ARPES II) is smaller than the corresponding DFT CESs (DFT $4-6$). Additionally, the  $|\mathbf{k}|$ values of DFT $4-6$ disperse more strongly with energy than for ARPES II, i.e., the experimental $E_{\rm peak,1}(\mathbf{k})$ is steeper in ($k_x$, $k_y$) direction. Again, we believe that these differences are caused by the additional order in the cubic Petrov phase, assumed for the DFT calculations.

\section*{Supplementary Note 2: Boltzmann relaxation model for an energy broadened M-shaped band}

\subsection*{Summary of basic equations}
The relation between electric field $\mathbf{E}$ and current density $\mathbf{j}$ in a two-dimensional solid (thin film) is:
\begin{equation}
	\begin{pmatrix} j_x \\ j_y \end{pmatrix} = \begin{pmatrix} \sigma_{xx} & \sigma_{xy} \\ \sigma_{yx} & \sigma_{yy} \end{pmatrix} \cdot \begin{pmatrix} E_x \\ E_y \end{pmatrix},
\end{equation}
where $\mathbf{\underline{\sigma}}$ is the $2\times 2$ conductivity matrix.\\
Within Boltzmann's relaxation model for a single spin degenerate band, one gets \cite{Ibach}:
\begin{eqnarray} \label{eq:19}
\sigma_{xx}&=&\frac{-e^{2}}{4\pi^3}  \iiint v^{2}_{x}(\mathbf{k})  \tau(\mathbf{k})  \frac{df_{0}(E_{\rm peak}(\mathbf{k}),T)}{dE_{\rm peak}} d^{3}\mathbf{k}\\ 
\sigma_{xy}&=&en_{\rm eff}/B.
\end{eqnarray}
Here, $v_{x}(\mathbf{k})$ is the group velocity in current direction $x$, $\tau(\mathbf{k})$ is the relaxation time, $f_{0}(E,T)$ is the Fermi distribution function, $B$ is the magnetic induction applied perpendicular to the ($x$,$y$) plane, and
\begin{equation}
\label{eq:neff}
n_{\rm eff} =\frac{2}{8\pi^3}\iiint_{{\rm Fermi\,volume}}d^3\mathbf{k},
\end{equation}
where Fermi volume is the Fermi volume of the spin-degenerate band, i.e., the volume within the BZ, which is enclosed by the corresponding Fermi surfaces.\\
Taking the band broadening into account, we have to replace the selected energies $E_{\rm peak}$ by the more general binding energy $E-E_{\rm F}$ and have to multiply the Fermi distribution function at $E$ with the normalized peak intensity at $E$ being $p_{\rm n}(E)$, i.e.:
\begin{equation}
\sigma_{xx}= 
\frac{-e^{2}}{4\pi^3}  \iiint v^{2}_{x}(\mathbf{k})  \tau(\mathbf{k}) p_{\rm n}(E) \frac{df_{0}(E(\mathbf{k}),T)}{dE} d^{3}\mathbf{k}. 
\end{equation}

For the sake of completeness, we finally add the standard derivation of the Drude result
at $T=0$ K, simplifying Supplementary Equation \ref{eq:19} to:
\begin{equation}
\sigma_{xx} = \frac{e^2}{4 \pi^3 \hbar}\sum_i \iint_{{\rm Fermi\, surface}_i} \frac{v^2_{x}(\mathbf{k}_\mathrm{F})}{|v_{\bot}(\mathbf{k}_\mathrm{F})|}\, \tau(\mathbf{k}_\mathrm{F}) \, d^2 \mathbf{k} \,
\label{eq:boltz}
\end{equation}
with the sum covering the possibly multiple Fermi surfaces $i$ of different spin degenerate bands, $\mathbf{k}_\mathrm{F}$ being the corresponding Fermi wave vector, and $v_{\bot}(\mathbf{k}_\mathrm{F})$ being the group velocity in the direction perpendicular to the Fermi surface. This ($T=0$~K)-approximation is reasonably valid as long as $\tau$, $v_{x}$, and $v_{\bot}$  vary negligibly within an energy interval of $k_{\rm B} T$ around $E_{\rm F}$ ($k_{\rm B}$: Boltzmann constant).\\
For a single, parabolic band with dispersion $E_{\rm peak}(\mathbf{k})=\hbar^2 |\mathbf{k}|^2/2m^\star$ and the assumption that $\tau(\mathbf{k}_{\rm F})$ is independent of $\mathbf{k}_{\rm F}$, one straightforwardly recovers the well-known Drude result:
\begin{equation}
\label{eq:Drude}
\ \sigma_{xx}=\frac{e^{2}n_{\rm eff}\tau}{m^{\star}}.
\end{equation}

\subsection*{Boltzmann model for the M-shaped valence band of GST-225}

Neglecting the disorder broadening, the upper valence band of GST-225 found by ARPES can be reasonably fitted by:
\begin{equation}
\label{eq:disp}
	E_{\rm peak}(\mathbf{k}) = E_{\rm peak, 0} - \frac{\hbar^2}{2m^\star} \left( k_\parallel-k_0 \right)^2 
\end{equation}
with $k_\parallel:=\sqrt{k_x^2 + k_y^2}$, i.e., an inverted, quadratic dispersion exhibiting rotational symmetry in the $(k_x,k_y)$ plane and no dispersion in $k_z$ direction. The cusp of the parabola is at $(E_{\rm peak, 0},k_0)$ and the curvature in radial in-plane direction is given by $m^\star$.\\

The resulting group velocities are:
\begin{eqnarray} \label{eq:vx}
	v_{x} &=& \frac{1}{\hbar} \frac{\partial E_{\rm peak}}{\partial k_x} = -\frac{\hbar}{m^\star} \frac{k_\parallel-k_0}{k_\parallel} k_x\\ 
   v_{\bot} &=&\frac{1}{\hbar} \frac{\partial E_{\rm peak}}{\partial k_\parallel} = -\frac{\hbar}{m^\star} \left( k_\parallel - k_0 \right),
\end{eqnarray}
where the latter exploits the cylindrical shape of the Fermi surface.\\

Defining $\varphi$ as the angle between $\mathbf{k}_\parallel= \left(k_x,k_y \right)$ and $k_x$, i.e., $k_x=k_\parallel \cos \varphi$, we get:
\begin{equation}
	\frac{v_{x}^2}{|v_{\bot}|} = \frac{\hbar}{m^\star} \frac{|k_\parallel-k_0|}{k_\parallel^2} k_x^2 = \frac{\hbar}{m^\star} \left| k_\parallel-k_0 \right| \cos^2 \varphi.
\end{equation}
Averaging over all angles $\varphi\in [0,2\pi)$ leads to:
\begin{equation}
	\overline{\frac{v_{x}^2}{v_{\bot}}} = \frac{\hbar}{2m^\star} \left| k_\parallel-k_0 \right|.
	\label{eq:vmb}
\end{equation}
Thus, Supplementary Equation \ref{eq:boltz} reads for $\overline{\tau}:=\tau(\mathbf{k}_{\rm F})$ independent of $\mathbf{k}_{\rm F}$:
\begin{equation}
\sigma_{xx} = \frac{\overline{\tau}e^2}{8 \pi^3 m^\star}\sum_{i=1}^2 \left| k_{{\rm F},xy,i}-k_0 \right| \iint_{{\rm Fermi\, surface}_i}\, d^2 \mathbf{k}
\end{equation}
with $k_{{\rm F},xy,i}$ being the radius of the $i.$ Fermi cylinder. Thus, we are left with the task to determine the area of the two cylindrical Fermi surfaces of the M-shaped band:
\begin{equation}
\iint_{{\rm Fermi\, surface}_i}\, d^2 \mathbf{k}= \frac{2\pi}{c}\cdot 2\pi k_{{\rm F},xy,i}
\end{equation}
with $c = 1.04$\,nm  being the extension of the unit cell of the disordered rock-salt phase along the stacking direction of the layers (Fig.\ 1a, c of main text).\cite{Nonaka2000}
Using $k_{{\rm F},xy,i}:=k_0\pm \Delta k$, i.e., exploiting the symmetric parabolicity of the band, we get:

\begin{eqnarray}
	\sigma_{xx} & = & \frac{\overline{\tau}e^2 }{2 \pi c m^\star} \left(\Delta k(k_0 + \Delta k) + \Delta k(k_0 - \Delta k) \right) \nonumber\\
	& = & \frac{\overline{\tau}e^2}{\pi c m^\star} k_0\Delta k
\end{eqnarray}\\

On the other hand, we can calculate $n_{\rm eff}$ (Supplementary Equation \ref{eq:neff}):
\begin{eqnarray}
\label{eq:neffM}
n_{\rm eff}&=& \frac{2}{8\pi^3}\frac{2\pi}{c} \pi((k_0+\Delta k)^2-(k_0-\Delta k)^2) \nonumber\\
&=& \frac{2}{\pi c}\Delta k\cdot k_0
\end{eqnarray}
leading to:
\begin{equation}
\sigma_{xx}= \frac{ e^2 n_{\rm eff}\overline{\tau}}{2 m^\star}.
\label{eq:31}
\end{equation}
Thus, the Drude result of Supplementary Equation \ref{eq:Drude} has simply to be divided by a factor of two.\\

Using Supplementary Equations \ref{eq:disp} and \ref{eq:neffM}, we can also deduce the relation between $n_{\rm eff}$ and the Fermi level $E_{\rm F}$:
\begin{equation}
E_{\rm F}-E_{\rm peak,0}=-\frac{\hbar} {8m^\star}\left( \frac{\pi c n_{\rm eff}}{k_0}\right)^2.
\end{equation}

\section*{Supplementary Note 3: Relation between scattering time and peak width}

Generally, one can argue that the mean free path $\lambda_{\rm MFP}$ of an electron sets the limit for its continuous wave-type propagation.
Hence, the electron wave function gets additionally structured on this length scale leading, e.g., to nodes at repulsive scatterers.
This can be approximated by $\lambda_{\rm MFP}\cdot|\Delta \mathbf{k}| \sim 1$ with a proportionality constant of order one, depending in detail on the
potential shape of the scatterers and the effective dimension \cite{Kramer1993,Evers2008}. Here, $|\Delta \mathbf{k}|$ describes the width of the peaks within the spectral function in momentum space.
Using $\lambda_{\rm MFP}=v_{\rm G} \cdot \tau$ with group velocity $v_{\rm G}$ and scattering time $\tau$ as well as $v_{\rm G}=\hbar^{-1} dE/d|\mathbf{k}|$, as valid for
an isotropic in-plane movement as largely present in GST-225, we get straightforwardly by Taylor expansion:
\begin{equation}
\Delta E \sim \frac{\hbar}{\tau}
\end{equation}

\section*{Supplementary Note 4: Crystal structures for the DFT calculations}

The cubic Petrov phase, as sketched in Fig.1d of the main text, is calculated using DFT in the generalized
gradient approximation \cite{Perdew1996} with the full-potential linearized augmented planewave method \cite{Wimmer1981}. The structure
is derived from the hexagonal Petrov phase by stacking three hexagonal unit cells and displacing them by $(-1/3, 1/3)$
with respect to each other in the [0001] plane. The resulting hexagonal unit cell has lattice parameters
$a = 4.257$~\AA~ and $c = 52.135$~\AA~ and atomic positions as indicated in Supplementary Table \ref{tab:pos}.

\begin{table}
\begin{tabular}{l|cr|cr|cr}
atom & $x,y$ & $z$ & $x,y$ & $z$& $x,y$ & $z$ \\
\hline
Te & B & $ 24.733$ & C & $  7.354$ & A & $-10.024$ \\
Ge & C & $ 23.903$ & A & $  5.931$ & B & $-11.447$ \\
Te & A & $ 21.183$ & B & $  3.804$ & C & $-13.574$ \\
Sb & B & $ 19.427$ & C & $  2.049$ & A & $-15.329$ \\
Te & C & $ 17.378$ & A & $  0.000$ & B & $-17.378$ \\
Sb & A & $ 15.329$ & B & $ -2.049$ & C & $-19.427$ \\
Te & B & $ 13.574$ & C & $ -3.804$ & A & $-21.183$ \\
Ge & C & $ 11.447$ & A & $ -5.931$ & B & $-23.903$ \\
Te & A & $ 10.024$ & B & $ -7.354$ & C & $-24.733$ \\
\end{tabular}
\caption{Atomic positions in the cubic Petrov phase used for the DFT calculations of the CECs.
The labels A, B, and C indicate in-plane ($x,y$) positions $(0, 0)$, $(-1/3, 1/3)$ and $(1/3, -1/3)$ in internal
coordinates, respectively. The numbers in column $z$ give the position in $z$-direction in \AA.} \label{tab:pos}
\end{table}

The muffin-tin radii, $R_i$, used in the calculations are $1.408$~\AA~ for Te and Ge and $1.445$~\AA~ for the Sb atoms.
The basis-set cutoff $R_{\rm min} k_{\rm max}$ was limited to $9$ and for the self-consistent calculations $26 \, \mathbf{k}$-points
were used in the irreducible Brillouin zone. For the plotting of the CECs, the reciprocal space was sampled
with $5210 \, \mathbf{k}$-points.
For the disordered surface models, a ZIP file provided additionally as Supporting Information contains the six structural models (in VASP CONTCAR format), exemplary input files (INCAR and KPOINTS) summarizing the parameters, as well as information on the particular pseudopotential files (POTCAR) employed for the computations.


\begin{thebibliography}{68}%
\makeatletter
\providecommand \@ifxundefined [1]{%
 \@ifx{#1\undefined}
}%
\providecommand \@ifnum [1]{%
 \ifnum #1\expandafter \@firstoftwo
 \else \expandafter \@secondoftwo
 \fi
}%
\providecommand \@ifx [1]{%
 \ifx #1\expandafter \@firstoftwo
 \else \expandafter \@secondoftwo
 \fi
}%
\providecommand \natexlab [1]{#1}%
\providecommand \enquote  [1]{``#1''}%
\providecommand \bibnamefont  [1]{#1}%
\providecommand \bibfnamefont [1]{#1}%
\providecommand \citenamefont [1]{#1}%
\providecommand \href@noop [0]{\@secondoftwo}%
\providecommand \href [0]{\begingroup \@sanitize@url \@href}%
\providecommand \@href[1]{\@@startlink{#1}\@@href}%
\providecommand \@@href[1]{\endgroup#1\@@endlink}%
\providecommand \@sanitize@url [0]{\catcode `\\12\catcode `\$12\catcode
  `\&12\catcode `\#12\catcode `\^12\catcode `\_12\catcode `\%12\relax}%
\providecommand \@@startlink[1]{}%
\providecommand \@@endlink[0]{}%
\providecommand \url  [0]{\begingroup\@sanitize@url \@url }%
\providecommand \@url [1]{\endgroup\@href {#1}{\urlprefix }}%
\providecommand \urlprefix  [0]{URL }%
\providecommand \Eprint [0]{\href }%
\providecommand \doibase [0]{http://dx.doi.org/}%
\providecommand \selectlanguage [0]{\@gobble}%
\providecommand \bibinfo  [0]{\@secondoftwo}%
\providecommand \bibfield  [0]{\@secondoftwo}%
\providecommand \translation [1]{[#1]}%
\providecommand \BibitemOpen [0]{}%
\providecommand \bibitemStop [0]{}%
\providecommand \bibitemNoStop [0]{.\EOS\space}%
\providecommand \EOS [0]{\spacefactor3000\relax}%
\providecommand \BibitemShut  [1]{\csname bibitem#1\endcsname}%
\let\auto@bib@innerbib\@empty
\bibitem [{\citenamefont {Wuttig}\ and\ \citenamefont
  {Yamada}(2007)}]{Wuttig2007}%
  \BibitemOpen
  \bibfield  {author} {\bibinfo {author} {\bibfnamefont {M.}~\bibnamefont
  {Wuttig}}\ and\ \bibinfo {author} {\bibfnamefont {N.}~\bibnamefont
  {Yamada}},\ }\href {\doibase 10.1038/nmat2009} {\bibfield  {journal}
  {\bibinfo  {journal} {Nat. Mater.}\ }\textbf {\bibinfo {volume} {6}},\
  \bibinfo {pages} {824} (\bibinfo {year} {2007})}\BibitemShut {NoStop}%
\bibitem [{\citenamefont {Wuttig}\ and\ \citenamefont
  {Raoux}(2012)}]{Wuttig2012}%
  \BibitemOpen
  \bibfield  {author} {\bibinfo {author} {\bibfnamefont {M.}~\bibnamefont
  {Wuttig}}\ and\ \bibinfo {author} {\bibfnamefont {S.}~\bibnamefont {Raoux}},\
  }\href {\doibase 10.1002/zaac.201200448} {\bibfield  {journal} {\bibinfo
  {journal} {Z. Anorg. Allg. Chem.}\ }\textbf {\bibinfo {volume} {638}},\
  \bibinfo {pages} {2455} (\bibinfo {year} {2012})}\BibitemShut {NoStop}%
\bibitem [{\citenamefont {Tominaga}\ \emph {et~al.}(2013)\citenamefont
  {Tominaga}, \citenamefont {Kolobov}, \citenamefont {Fons}, \citenamefont
  {Nakano},\ and\ \citenamefont {Murakami}}]{Tominaga2014}%
  \BibitemOpen
  \bibfield  {author} {\bibinfo {author} {\bibfnamefont {J.}~\bibnamefont
  {Tominaga}}, \bibinfo {author} {\bibfnamefont {A.~V.}\ \bibnamefont
  {Kolobov}}, \bibinfo {author} {\bibfnamefont {P.}~\bibnamefont {Fons}},
  \bibinfo {author} {\bibfnamefont {T.}~\bibnamefont {Nakano}}, \ and\ \bibinfo
  {author} {\bibfnamefont {S.}~\bibnamefont {Murakami}},\ }\href {\doibase
  10.1002/admi.201300027} {\bibfield  {journal} {\bibinfo  {journal} {Adv.
  Mater. Interfaces}\ }\textbf {\bibinfo {volume} {1}},\ \bibinfo {pages}
  {1300027} (\bibinfo {year} {2013})}\BibitemShut {NoStop}%
\bibitem [{\citenamefont {Deringer}\ \emph {et~al.}(2015)\citenamefont
  {Deringer}, \citenamefont {Dronskowski},\ and\ \citenamefont
  {Wuttig}}]{Deringer2015}%
  \BibitemOpen
  \bibfield  {author} {\bibinfo {author} {\bibfnamefont {V.~L.}\ \bibnamefont
  {Deringer}}, \bibinfo {author} {\bibfnamefont {R.}~\bibnamefont
  {Dronskowski}}, \ and\ \bibinfo {author} {\bibfnamefont {M.}~\bibnamefont
  {Wuttig}},\ }\href {\doibase 10.1002/adfm.201500826} {\bibfield  {journal}
  {\bibinfo  {journal} {Adv. Funct. Mater.}\ }\textbf {\bibinfo {volume}
  {25}},\ \bibinfo {pages} {6343} (\bibinfo {year} {2015})}\BibitemShut
  {NoStop}%
\bibitem [{\citenamefont {Ovshinsky}(1968)}]{Ovshinsky1968}%
  \BibitemOpen
  \bibfield  {author} {\bibinfo {author} {\bibfnamefont {S.~R.}\ \bibnamefont
  {Ovshinsky}},\ }\href {\doibase 10.1103/physrevlett.21.1450} {\bibfield
  {journal} {\bibinfo  {journal} {Phys. Rev. Lett.}\ }\textbf {\bibinfo
  {volume} {21}},\ \bibinfo {pages} {1450} (\bibinfo {year}
  {1968})}\BibitemShut {NoStop}%
\bibitem [{\citenamefont {Yamada}\ \emph {et~al.}(1987)\citenamefont {Yamada},
  \citenamefont {Ohno}, \citenamefont {Akahira}, \citenamefont {Nishiuchi},
  \citenamefont {Nagata},\ and\ \citenamefont {Takao}}]{Yamada1987}%
  \BibitemOpen
  \bibfield  {author} {\bibinfo {author} {\bibfnamefont {N.}~\bibnamefont
  {Yamada}}, \bibinfo {author} {\bibfnamefont {E.}~\bibnamefont {Ohno}},
  \bibinfo {author} {\bibfnamefont {N.}~\bibnamefont {Akahira}}, \bibinfo
  {author} {\bibfnamefont {K.}~\bibnamefont {Nishiuchi}}, \bibinfo {author}
  {\bibfnamefont {K.}~\bibnamefont {Nagata}}, \ and\ \bibinfo {author}
  {\bibfnamefont {M.}~\bibnamefont {Takao}},\ }\href {\doibase
  10.7567/jjaps.26s4.61} {\bibfield  {journal} {\bibinfo  {journal} {Jpn. J.
  Appl. Phys.}\ }\textbf {\bibinfo {volume} {26}},\ \bibinfo {pages} {61}
  (\bibinfo {year} {1987})}\BibitemShut {NoStop}%
\bibitem [{\citenamefont {Yamada}\ \emph {et~al.}(1991)\citenamefont {Yamada},
  \citenamefont {Ohno}, \citenamefont {Nishiuchi}, \citenamefont {Akahira},\
  and\ \citenamefont {Takao}}]{Yamada1991}%
  \BibitemOpen
  \bibfield  {author} {\bibinfo {author} {\bibfnamefont {N.}~\bibnamefont
  {Yamada}}, \bibinfo {author} {\bibfnamefont {E.}~\bibnamefont {Ohno}},
  \bibinfo {author} {\bibfnamefont {K.}~\bibnamefont {Nishiuchi}}, \bibinfo
  {author} {\bibfnamefont {N.}~\bibnamefont {Akahira}}, \ and\ \bibinfo
  {author} {\bibfnamefont {M.}~\bibnamefont {Takao}},\ }\href {\doibase
  10.1063/1.348620} {\bibfield  {journal} {\bibinfo  {journal} {J. Appl.
  Phys.}\ }\textbf {\bibinfo {volume} {69}},\ \bibinfo {pages} {2849} (\bibinfo
  {year} {1991})}\BibitemShut {NoStop}%
\bibitem [{\citenamefont {Loke}\ \emph {et~al.}(2012)\citenamefont {Loke},
  \citenamefont {Lee}, \citenamefont {Wang}, \citenamefont {Shi}, \citenamefont
  {Zhao}, \citenamefont {Yeo}, \citenamefont {Chong},\ and\ \citenamefont
  {Elliott}}]{Loke2012}%
  \BibitemOpen
  \bibfield  {author} {\bibinfo {author} {\bibfnamefont {D.}~\bibnamefont
  {Loke}}, \bibinfo {author} {\bibfnamefont {T.~H.}\ \bibnamefont {Lee}},
  \bibinfo {author} {\bibfnamefont {W.~J.}\ \bibnamefont {Wang}}, \bibinfo
  {author} {\bibfnamefont {L.~P.}\ \bibnamefont {Shi}}, \bibinfo {author}
  {\bibfnamefont {R.}~\bibnamefont {Zhao}}, \bibinfo {author} {\bibfnamefont
  {Y.~C.}\ \bibnamefont {Yeo}}, \bibinfo {author} {\bibfnamefont {T.~C.}\
  \bibnamefont {Chong}}, \ and\ \bibinfo {author} {\bibfnamefont {S.~R.}\
  \bibnamefont {Elliott}},\ }\href {\doibase 10.1126/science.1221561}
  {\bibfield  {journal} {\bibinfo  {journal} {Science}\ }\textbf {\bibinfo
  {volume} {336}},\ \bibinfo {pages} {1566} (\bibinfo {year}
  {2012})}\BibitemShut {NoStop}%
\bibitem [{\citenamefont {Xiong}\ \emph {et~al.}(2011)\citenamefont {Xiong},
  \citenamefont {Liao}, \citenamefont {Estrada},\ and\ \citenamefont
  {Pop.}}]{Xiong2011}%
  \BibitemOpen
  \bibfield  {author} {\bibinfo {author} {\bibfnamefont {F.}~\bibnamefont
  {Xiong}}, \bibinfo {author} {\bibfnamefont {A.~D.}\ \bibnamefont {Liao}},
  \bibinfo {author} {\bibfnamefont {D.}~\bibnamefont {Estrada}}, \ and\
  \bibinfo {author} {\bibfnamefont {E.}~\bibnamefont {Pop.}},\ }\href {\doibase
  10.1126/science.1201938} {\bibfield  {journal} {\bibinfo  {journal}
  {Science}\ }\textbf {\bibinfo {volume} {332}},\ \bibinfo {pages} {568}
  (\bibinfo {year} {2011})}\BibitemShut {NoStop}%
\bibitem [{\citenamefont {Park}\ \emph {et~al.}(2007)\citenamefont {Park},
  \citenamefont {Park}, \citenamefont {Baik}, \citenamefont {Lee},
  \citenamefont {Jeong},\ and\ \citenamefont {Kim}}]{Park2007}%
  \BibitemOpen
  \bibfield  {author} {\bibinfo {author} {\bibfnamefont {J.-B.}\ \bibnamefont
  {Park}}, \bibinfo {author} {\bibfnamefont {G.-S.}\ \bibnamefont {Park}},
  \bibinfo {author} {\bibfnamefont {H.-S.}\ \bibnamefont {Baik}}, \bibinfo
  {author} {\bibfnamefont {J.-H.}\ \bibnamefont {Lee}}, \bibinfo {author}
  {\bibfnamefont {H.}~\bibnamefont {Jeong}}, \ and\ \bibinfo {author}
  {\bibfnamefont {K.}~\bibnamefont {Kim}},\ }\href {\doibase 10.1149/1.2409482}
  {\bibfield  {journal} {\bibinfo  {journal} {J. Electrochem. Soc.}\ }\textbf
  {\bibinfo {volume} {154}},\ \bibinfo {pages} {H139} (\bibinfo {year}
  {2007})}\BibitemShut {NoStop}%
\bibitem [{\citenamefont {Matsunaga}\ and\ \citenamefont
  {Yamada}(2004)}]{Matsunaga2004}%
  \BibitemOpen
  \bibfield  {author} {\bibinfo {author} {\bibfnamefont {T.}~\bibnamefont
  {Matsunaga}}\ and\ \bibinfo {author} {\bibfnamefont {N.}~\bibnamefont
  {Yamada}},\ }\href {\doibase 10.1103/physrevb.69.104111} {\bibfield
  {journal} {\bibinfo  {journal} {Phys. Rev. B}\ }\textbf {\bibinfo {volume}
  {69}},\ \bibinfo {pages} {104111} (\bibinfo {year} {2004})}\BibitemShut
  {NoStop}%
\bibitem [{\citenamefont {Matsunaga}\ \emph {et~al.}(2006)\citenamefont
  {Matsunaga}, \citenamefont {Kojima}, \citenamefont {Yamada}, \citenamefont
  {Kifune}, \citenamefont {Kubota}, \citenamefont {Tabata},\ and\ \citenamefont
  {Takata}}]{Matsunaga2006}%
  \BibitemOpen
  \bibfield  {author} {\bibinfo {author} {\bibfnamefont {T.}~\bibnamefont
  {Matsunaga}}, \bibinfo {author} {\bibfnamefont {R.}~\bibnamefont {Kojima}},
  \bibinfo {author} {\bibfnamefont {N.}~\bibnamefont {Yamada}}, \bibinfo
  {author} {\bibfnamefont {K.}~\bibnamefont {Kifune}}, \bibinfo {author}
  {\bibfnamefont {Y.}~\bibnamefont {Kubota}}, \bibinfo {author} {\bibfnamefont
  {Y.}~\bibnamefont {Tabata}}, \ and\ \bibinfo {author} {\bibfnamefont
  {M.}~\bibnamefont {Takata}},\ }\href {\doibase 10.1021/ic051677w} {\bibfield
  {journal} {\bibinfo  {journal} {Inorg. Chem.}\ }\textbf {\bibinfo {volume}
  {45}},\ \bibinfo {pages} {2235} (\bibinfo {year} {2006})}\BibitemShut
  {NoStop}%
\bibitem [{\citenamefont {Matsunaga}\ \emph {et~al.}(2008)\citenamefont
  {Matsunaga}, \citenamefont {Morita}, \citenamefont {Kojima}, \citenamefont
  {Yamada}, \citenamefont {Kifune}, \citenamefont {Kubota}, \citenamefont
  {Tabata}, \citenamefont {Kim}, \citenamefont {Kobata}, \citenamefont
  {Ikenaga},\ and\ \citenamefont {Kobayashi}}]{Matsunaga2008}%
  \BibitemOpen
  \bibfield  {author} {\bibinfo {author} {\bibfnamefont {T.}~\bibnamefont
  {Matsunaga}}, \bibinfo {author} {\bibfnamefont {H.}~\bibnamefont {Morita}},
  \bibinfo {author} {\bibfnamefont {R.}~\bibnamefont {Kojima}}, \bibinfo
  {author} {\bibfnamefont {N.}~\bibnamefont {Yamada}}, \bibinfo {author}
  {\bibfnamefont {K.}~\bibnamefont {Kifune}}, \bibinfo {author} {\bibfnamefont
  {Y.}~\bibnamefont {Kubota}}, \bibinfo {author} {\bibfnamefont
  {Y.}~\bibnamefont {Tabata}}, \bibinfo {author} {\bibfnamefont {J.-J.}\
  \bibnamefont {Kim}}, \bibinfo {author} {\bibfnamefont {M.}~\bibnamefont
  {Kobata}}, \bibinfo {author} {\bibfnamefont {E.}~\bibnamefont {Ikenaga}}, \
  and\ \bibinfo {author} {\bibfnamefont {K.}~\bibnamefont {Kobayashi}},\ }\href
  {\doibase 10.1063/1.2901187} {\bibfield  {journal} {\bibinfo  {journal} {J.
  Appl. Phys.}\ }\textbf {\bibinfo {volume} {103}},\ \bibinfo {pages} {093511}
  (\bibinfo {year} {2008})}\BibitemShut {NoStop}%
\bibitem [{\citenamefont {Silva}\ \emph {et~al.}(2008)\citenamefont {Silva},
  \citenamefont {Walsh},\ and\ \citenamefont {Lee}}]{Silva2008}%
  \BibitemOpen
  \bibfield  {author} {\bibinfo {author} {\bibfnamefont {J.~L. F.~D.}\
  \bibnamefont {Silva}}, \bibinfo {author} {\bibfnamefont {A.}~\bibnamefont
  {Walsh}}, \ and\ \bibinfo {author} {\bibfnamefont {H.}~\bibnamefont {Lee}},\
  }\href {\doibase 10.1103/physrevb.78.224111} {\bibfield  {journal} {\bibinfo
  {journal} {Phys. Rev. B}\ }\textbf {\bibinfo {volume} {78}},\ \bibinfo
  {pages} {224111} (\bibinfo {year} {2008})}\BibitemShut {NoStop}%
\bibitem [{\citenamefont {Petrov}\ \emph {et~al.}(1968)\citenamefont {Petrov},
  \citenamefont {Imamov},\ and\ \citenamefont {Pinsker}}]{Petrov1968}%
  \BibitemOpen
  \bibfield  {author} {\bibinfo {author} {\bibfnamefont {I.~I.}\ \bibnamefont
  {Petrov}}, \bibinfo {author} {\bibfnamefont {R.~M.}\ \bibnamefont {Imamov}},
  \ and\ \bibinfo {author} {\bibfnamefont {Z.~G.}\ \bibnamefont {Pinsker}},\
  }\href@noop {} {\bibfield  {journal} {\bibinfo  {journal} {Sov. Phys.
  Crystallogr.}\ }\textbf {\bibinfo {volume} {13}},\ \bibinfo {pages} {339}
  (\bibinfo {year} {1968})}\BibitemShut {NoStop}%
\bibitem [{\citenamefont {Kooi}\ and\ \citenamefont {Hosson}(2002)}]{Kooi2002}%
  \BibitemOpen
  \bibfield  {author} {\bibinfo {author} {\bibfnamefont {B.~J.}\ \bibnamefont
  {Kooi}}\ and\ \bibinfo {author} {\bibfnamefont {J.~T. M.~D.}\ \bibnamefont
  {Hosson}},\ }\href {\doibase 10.1063/1.1502915} {\bibfield  {journal}
  {\bibinfo  {journal} {J. Appl. Phys.}\ }\textbf {\bibinfo {volume} {92}},\
  \bibinfo {pages} {3584} (\bibinfo {year} {2002})}\BibitemShut {NoStop}%
\bibitem [{\citenamefont {Siegrist}\ \emph {et~al.}(2011)\citenamefont
  {Siegrist}, \citenamefont {Jost}, \citenamefont {Volker}, \citenamefont
  {Woda}, \citenamefont {Merkelbach}, \citenamefont {Schlockermann},\ and\
  \citenamefont {Wuttig}}]{Siegrist2011}%
  \BibitemOpen
  \bibfield  {author} {\bibinfo {author} {\bibfnamefont {T.}~\bibnamefont
  {Siegrist}}, \bibinfo {author} {\bibfnamefont {P.}~\bibnamefont {Jost}},
  \bibinfo {author} {\bibfnamefont {H.}~\bibnamefont {Volker}}, \bibinfo
  {author} {\bibfnamefont {M.}~\bibnamefont {Woda}}, \bibinfo {author}
  {\bibfnamefont {P.}~\bibnamefont {Merkelbach}}, \bibinfo {author}
  {\bibfnamefont {C.}~\bibnamefont {Schlockermann}}, \ and\ \bibinfo {author}
  {\bibfnamefont {M.}~\bibnamefont {Wuttig}},\ }\href {\doibase
  10.1038/nmat2934} {\bibfield  {journal} {\bibinfo  {journal} {Nat. Mater.}\
  }\textbf {\bibinfo {volume} {10}},\ \bibinfo {pages} {202} (\bibinfo {year}
  {2011})}\BibitemShut {NoStop}%
\bibitem [{\citenamefont {Schneider}\ \emph {et~al.}(2012)\citenamefont
  {Schneider}, \citenamefont {Biquard}, \citenamefont {Stiewe}, \citenamefont
  {Schr{\"o}der}, \citenamefont {Urban},\ and\ \citenamefont
  {Oeckler}}]{Schneider2012}%
  \BibitemOpen
  \bibfield  {author} {\bibinfo {author} {\bibfnamefont {M.~N.}\ \bibnamefont
  {Schneider}}, \bibinfo {author} {\bibfnamefont {X.}~\bibnamefont {Biquard}},
  \bibinfo {author} {\bibfnamefont {C.}~\bibnamefont {Stiewe}}, \bibinfo
  {author} {\bibfnamefont {T.}~\bibnamefont {Schr{\"o}der}}, \bibinfo {author}
  {\bibfnamefont {P.}~\bibnamefont {Urban}}, \ and\ \bibinfo {author}
  {\bibfnamefont {O.}~\bibnamefont {Oeckler}},\ }\href {\doibase
  10.1039/C1CC15808B} {\bibfield  {journal} {\bibinfo  {journal} {Chem.
  Commun.}\ }\textbf {\bibinfo {volume} {48}},\ \bibinfo {pages} {2192}
  (\bibinfo {year} {2012})}\BibitemShut {NoStop}%
\bibitem [{\citenamefont {Zhang}\ \emph {et~al.}(2012)\citenamefont {Zhang},
  \citenamefont {Thiess}, \citenamefont {Zalden}, \citenamefont {Zeller},
  \citenamefont {Dederichs}, \citenamefont {Raty}, \citenamefont {Wuttig},
  \citenamefont {Bl{\"u}gel},\ and\ \citenamefont {Mazzarello}}]{Zhang2012}%
  \BibitemOpen
  \bibfield  {author} {\bibinfo {author} {\bibfnamefont {W.}~\bibnamefont
  {Zhang}}, \bibinfo {author} {\bibfnamefont {A.}~\bibnamefont {Thiess}},
  \bibinfo {author} {\bibfnamefont {P.}~\bibnamefont {Zalden}}, \bibinfo
  {author} {\bibfnamefont {R.}~\bibnamefont {Zeller}}, \bibinfo {author}
  {\bibfnamefont {P.~H.}\ \bibnamefont {Dederichs}}, \bibinfo {author}
  {\bibfnamefont {J.-Y.}\ \bibnamefont {Raty}}, \bibinfo {author}
  {\bibfnamefont {M.}~\bibnamefont {Wuttig}}, \bibinfo {author} {\bibfnamefont
  {S.}~\bibnamefont {Bl{\"u}gel}}, \ and\ \bibinfo {author} {\bibfnamefont
  {R.}~\bibnamefont {Mazzarello}},\ }\href {\doibase 10.1038/nmat3456}
  {\bibfield  {journal} {\bibinfo  {journal} {Nat. Mater.}\ }\textbf {\bibinfo
  {volume} {11}},\ \bibinfo {pages} {952} (\bibinfo {year} {2012})}\BibitemShut
  {NoStop}%
\bibitem [{\citenamefont {Bragaglia}\ \emph
  {et~al.}(2016{\natexlab{a}})\citenamefont {Bragaglia}, \citenamefont
  {Arciprete}, \citenamefont {Zhang}, \citenamefont {Mio}, \citenamefont
  {Zallo}, \citenamefont {Perumal}, \citenamefont {Giussani}, \citenamefont
  {Cecchi}, \citenamefont {Boschker}, \citenamefont {Riechert}, \citenamefont
  {Privitera}, \citenamefont {Rimini}, \citenamefont {Mazzarello},\ and\
  \citenamefont {Calarco}}]{Bragaglia2016}%
  \BibitemOpen
  \bibfield  {author} {\bibinfo {author} {\bibfnamefont {V.}~\bibnamefont
  {Bragaglia}}, \bibinfo {author} {\bibfnamefont {F.}~\bibnamefont
  {Arciprete}}, \bibinfo {author} {\bibfnamefont {W.}~\bibnamefont {Zhang}},
  \bibinfo {author} {\bibfnamefont {A.~M.}\ \bibnamefont {Mio}}, \bibinfo
  {author} {\bibfnamefont {E.}~\bibnamefont {Zallo}}, \bibinfo {author}
  {\bibfnamefont {K.}~\bibnamefont {Perumal}}, \bibinfo {author} {\bibfnamefont
  {A.}~\bibnamefont {Giussani}}, \bibinfo {author} {\bibfnamefont
  {S.}~\bibnamefont {Cecchi}}, \bibinfo {author} {\bibfnamefont {J.~E.}\
  \bibnamefont {Boschker}}, \bibinfo {author} {\bibfnamefont {H.}~\bibnamefont
  {Riechert}}, \bibinfo {author} {\bibfnamefont {S.}~\bibnamefont {Privitera}},
  \bibinfo {author} {\bibfnamefont {E.}~\bibnamefont {Rimini}}, \bibinfo
  {author} {\bibfnamefont {R.}~\bibnamefont {Mazzarello}}, \ and\ \bibinfo
  {author} {\bibfnamefont {R.}~\bibnamefont {Calarco}},\ }\href {\doibase
  10.1038/srep23843} {\bibfield  {journal} {\bibinfo  {journal} {Sci. Rep.}\
  }\textbf {\bibinfo {volume} {6}},\ \bibinfo {pages} {23843} (\bibinfo {year}
  {2016}{\natexlab{a}})}\BibitemShut {NoStop}%
\bibitem [{\citenamefont {Subramaniam}\ \emph {et~al.}(2009)\citenamefont
  {Subramaniam}, \citenamefont {Pauly}, \citenamefont {Liebmann}, \citenamefont
  {Woda}, \citenamefont {Rausch}, \citenamefont {Merkelbach}, \citenamefont
  {Wuttig},\ and\ \citenamefont {Morgenstern}}]{Subramaniam2009}%
  \BibitemOpen
  \bibfield  {author} {\bibinfo {author} {\bibfnamefont {D.}~\bibnamefont
  {Subramaniam}}, \bibinfo {author} {\bibfnamefont {C.}~\bibnamefont {Pauly}},
  \bibinfo {author} {\bibfnamefont {M.}~\bibnamefont {Liebmann}}, \bibinfo
  {author} {\bibfnamefont {M.}~\bibnamefont {Woda}}, \bibinfo {author}
  {\bibfnamefont {P.}~\bibnamefont {Rausch}}, \bibinfo {author} {\bibfnamefont
  {P.}~\bibnamefont {Merkelbach}}, \bibinfo {author} {\bibfnamefont
  {M.}~\bibnamefont {Wuttig}}, \ and\ \bibinfo {author} {\bibfnamefont
  {M.}~\bibnamefont {Morgenstern}},\ }\href {\doibase 10.1063/1.3211991}
  {\bibfield  {journal} {\bibinfo  {journal} {Appl. Phys. Lett.}\ }\textbf
  {\bibinfo {volume} {95}},\ \bibinfo {pages} {103110} (\bibinfo {year}
  {2009})}\BibitemShut {NoStop}%
\bibitem [{\citenamefont {Volker}(2013)}]{Volker}%
  \BibitemOpen
  \bibfield  {author} {\bibinfo {author} {\bibfnamefont {H.}~\bibnamefont
  {Volker}},\ }\emph {\bibinfo {title} {Disorder and electrical transport in
  phase-change materials}},\ \href@noop {} {Ph.D. thesis},\ \bibinfo  {school}
  {RWTH Aachen University} (\bibinfo {year} {2013})\BibitemShut {NoStop}%
\bibitem [{\citenamefont {Katmis}\ \emph {et~al.}(2011)\citenamefont {Katmis},
  \citenamefont {Calarco}, \citenamefont {Perumal}, \citenamefont {Rodenbach},
  \citenamefont {Giussani}, \citenamefont {M.Hanke}, \citenamefont
  {Proessdorf}, \citenamefont {Trampert}, \citenamefont {Grosse}, \citenamefont
  {Shayduk}, \citenamefont {Campion}, \citenamefont {Braun},\ and\
  \citenamefont {Riechert}}]{Katmis2011}%
  \BibitemOpen
  \bibfield  {author} {\bibinfo {author} {\bibfnamefont {F.}~\bibnamefont
  {Katmis}}, \bibinfo {author} {\bibfnamefont {R.}~\bibnamefont {Calarco}},
  \bibinfo {author} {\bibfnamefont {K.}~\bibnamefont {Perumal}}, \bibinfo
  {author} {\bibfnamefont {P.}~\bibnamefont {Rodenbach}}, \bibinfo {author}
  {\bibfnamefont {A.}~\bibnamefont {Giussani}}, \bibinfo {author} {\bibnamefont
  {M.Hanke}}, \bibinfo {author} {\bibfnamefont {A.}~\bibnamefont {Proessdorf}},
  \bibinfo {author} {\bibfnamefont {A.}~\bibnamefont {Trampert}}, \bibinfo
  {author} {\bibfnamefont {F.}~\bibnamefont {Grosse}}, \bibinfo {author}
  {\bibfnamefont {R.}~\bibnamefont {Shayduk}}, \bibinfo {author} {\bibfnamefont
  {R.}~\bibnamefont {Campion}}, \bibinfo {author} {\bibfnamefont
  {W.}~\bibnamefont {Braun}}, \ and\ \bibinfo {author} {\bibfnamefont
  {H.}~\bibnamefont {Riechert}},\ }\href {\doibase 10.1021/cg200857x}
  {\bibfield  {journal} {\bibinfo  {journal} {Cryst. Growth Des.}\ }\textbf
  {\bibinfo {volume} {11}},\ \bibinfo {pages} {4606} (\bibinfo {year}
  {2011})}\BibitemShut {NoStop}%
\bibitem [{\citenamefont {Rodenbach}\ \emph {et~al.}(2012)\citenamefont
  {Rodenbach}, \citenamefont {Calarco}, \citenamefont {Perumal}, \citenamefont
  {Katmis}, \citenamefont {Hanke}, \citenamefont {Proessdorf}, \citenamefont
  {Braun}, \citenamefont {Giussani}, \citenamefont {Trampert}, \citenamefont
  {Riechert}, \citenamefont {Fons},\ and\ \citenamefont
  {Kolobov}}]{Rodenbach2012}%
  \BibitemOpen
  \bibfield  {author} {\bibinfo {author} {\bibfnamefont {P.}~\bibnamefont
  {Rodenbach}}, \bibinfo {author} {\bibfnamefont {R.}~\bibnamefont {Calarco}},
  \bibinfo {author} {\bibfnamefont {K.}~\bibnamefont {Perumal}}, \bibinfo
  {author} {\bibfnamefont {F.}~\bibnamefont {Katmis}}, \bibinfo {author}
  {\bibfnamefont {M.}~\bibnamefont {Hanke}}, \bibinfo {author} {\bibfnamefont
  {A.}~\bibnamefont {Proessdorf}}, \bibinfo {author} {\bibfnamefont
  {W.}~\bibnamefont {Braun}}, \bibinfo {author} {\bibfnamefont
  {A.}~\bibnamefont {Giussani}}, \bibinfo {author} {\bibfnamefont
  {A.}~\bibnamefont {Trampert}}, \bibinfo {author} {\bibfnamefont
  {H.}~\bibnamefont {Riechert}}, \bibinfo {author} {\bibfnamefont
  {P.}~\bibnamefont {Fons}}, \ and\ \bibinfo {author} {\bibfnamefont {A.~V.}\
  \bibnamefont {Kolobov}},\ }\href {\doibase 10.1002/pssr.201206387} {\bibfield
   {journal} {\bibinfo  {journal} {Phys. Status Solidi-R}\ }\textbf {\bibinfo
  {volume} {6}},\ \bibinfo {pages} {415} (\bibinfo {year} {2012})}\BibitemShut
  {NoStop}%
\bibitem [{\citenamefont {Bragaglia}\ \emph {et~al.}(2014)\citenamefont
  {Bragaglia}, \citenamefont {Jenichen}, \citenamefont {Giussani},
  \citenamefont {Riechert},\ and\ \citenamefont {Calarco}}]{Bragaglia2014}%
  \BibitemOpen
  \bibfield  {author} {\bibinfo {author} {\bibfnamefont {V.}~\bibnamefont
  {Bragaglia}}, \bibinfo {author} {\bibfnamefont {B.}~\bibnamefont {Jenichen}},
  \bibinfo {author} {\bibfnamefont {A.}~\bibnamefont {Giussani}}, \bibinfo
  {author} {\bibfnamefont {K.~P.~H.}\ \bibnamefont {Riechert}}, \ and\ \bibinfo
  {author} {\bibfnamefont {R.}~\bibnamefont {Calarco}},\ }\href {\doibase
  10.1063/1.4892394} {\bibfield  {journal} {\bibinfo  {journal} {J. Appl.
  Phys.}\ }\textbf {\bibinfo {volume} {116}},\ \bibinfo {pages} {054913}
  (\bibinfo {year} {2014})}\BibitemShut {NoStop}%
\bibitem [{\citenamefont {Cecchi}\ \emph {et~al.}(2017)\citenamefont {Cecchi},
  \citenamefont {Zallo}, \citenamefont {Momand}, \citenamefont {Wang},
  \citenamefont {Kooi}, \citenamefont {Verheijen},\ and\ \citenamefont
  {Calarco}}]{Cecchi2017}%
  \BibitemOpen
  \bibfield  {author} {\bibinfo {author} {\bibfnamefont {S.}~\bibnamefont
  {Cecchi}}, \bibinfo {author} {\bibfnamefont {E.}~\bibnamefont {Zallo}},
  \bibinfo {author} {\bibfnamefont {J.}~\bibnamefont {Momand}}, \bibinfo
  {author} {\bibfnamefont {R.}~\bibnamefont {Wang}}, \bibinfo {author}
  {\bibfnamefont {B.}~\bibnamefont {Kooi}}, \bibinfo {author} {\bibfnamefont
  {M.}~\bibnamefont {Verheijen}}, \ and\ \bibinfo {author} {\bibfnamefont
  {R.}~\bibnamefont {Calarco}},\ }\href {\doibase 10.1063/1.4976828} {\bibfield
   {journal} {\bibinfo  {journal} {APL Mater.}\ }\textbf {\bibinfo {volume}
  {5}},\ \bibinfo {pages} {026107} (\bibinfo {year} {2017})}\BibitemShut
  {NoStop}%
\bibitem [{\citenamefont {Mitrofanov}\ \emph {et~al.}(2016)\citenamefont
  {Mitrofanov}, \citenamefont {Fons}, \citenamefont {Makino}, \citenamefont
  {Terashima}, \citenamefont {Shimada}, \citenamefont {Kolobov}, \citenamefont
  {Tominaga}, \citenamefont {Bragaglia}, \citenamefont {Giussani},
  \citenamefont {Calarco}, \citenamefont {Riechert}, \citenamefont {Sato},
  \citenamefont {Katayama}, \citenamefont {Ogawa}, \citenamefont {Togashi},
  \citenamefont {Yabashi}, \citenamefont {Wall}, \citenamefont {Brewe},\ and\
  \citenamefont {Hase}}]{Mitrofanov2016}%
  \BibitemOpen
  \bibfield  {author} {\bibinfo {author} {\bibfnamefont {K.~V.}\ \bibnamefont
  {Mitrofanov}}, \bibinfo {author} {\bibfnamefont {P.}~\bibnamefont {Fons}},
  \bibinfo {author} {\bibfnamefont {K.}~\bibnamefont {Makino}}, \bibinfo
  {author} {\bibfnamefont {R.}~\bibnamefont {Terashima}}, \bibinfo {author}
  {\bibfnamefont {T.}~\bibnamefont {Shimada}}, \bibinfo {author} {\bibfnamefont
  {A.~V.}\ \bibnamefont {Kolobov}}, \bibinfo {author} {\bibfnamefont
  {J.}~\bibnamefont {Tominaga}}, \bibinfo {author} {\bibfnamefont
  {V.}~\bibnamefont {Bragaglia}}, \bibinfo {author} {\bibfnamefont
  {A.}~\bibnamefont {Giussani}}, \bibinfo {author} {\bibfnamefont
  {R.}~\bibnamefont {Calarco}}, \bibinfo {author} {\bibfnamefont
  {H.}~\bibnamefont {Riechert}}, \bibinfo {author} {\bibfnamefont
  {T.}~\bibnamefont {Sato}}, \bibinfo {author} {\bibfnamefont {T.}~\bibnamefont
  {Katayama}}, \bibinfo {author} {\bibfnamefont {K.}~\bibnamefont {Ogawa}},
  \bibinfo {author} {\bibfnamefont {T.}~\bibnamefont {Togashi}}, \bibinfo
  {author} {\bibfnamefont {M.}~\bibnamefont {Yabashi}}, \bibinfo {author}
  {\bibfnamefont {S.}~\bibnamefont {Wall}}, \bibinfo {author} {\bibfnamefont
  {D.}~\bibnamefont {Brewe}}, \ and\ \bibinfo {author} {\bibfnamefont
  {M.}~\bibnamefont {Hase}},\ }\href {\doibase 10.1038/srep20633} {\bibfield
  {journal} {\bibinfo  {journal} {Sci. Rep.}\ }\textbf {\bibinfo {volume}
  {6}},\ \bibinfo {pages} {20633} (\bibinfo {year} {2016})}\BibitemShut
  {NoStop}%
\bibitem [{\citenamefont {Bragaglia}\ \emph
  {et~al.}(2016{\natexlab{b}})\citenamefont {Bragaglia}, \citenamefont
  {Holldack}, \citenamefont {Boschker}, \citenamefont {Arciprete},
  \citenamefont {Zallo}, \citenamefont {Flissikowski},\ and\ \citenamefont
  {Calarco}}]{Bragaglia2016a}%
  \BibitemOpen
  \bibfield  {author} {\bibinfo {author} {\bibfnamefont {V.}~\bibnamefont
  {Bragaglia}}, \bibinfo {author} {\bibfnamefont {K.}~\bibnamefont {Holldack}},
  \bibinfo {author} {\bibfnamefont {J.~E.}\ \bibnamefont {Boschker}}, \bibinfo
  {author} {\bibfnamefont {F.}~\bibnamefont {Arciprete}}, \bibinfo {author}
  {\bibfnamefont {E.}~\bibnamefont {Zallo}}, \bibinfo {author} {\bibfnamefont
  {T.}~\bibnamefont {Flissikowski}}, \ and\ \bibinfo {author} {\bibfnamefont
  {R.}~\bibnamefont {Calarco}},\ }\href {\doibase 10.1038/srep28560} {\bibfield
   {journal} {\bibinfo  {journal} {Sci. Rep.}\ }\textbf {\bibinfo {volume}
  {6}},\ \bibinfo {pages} {28560} (\bibinfo {year}
  {2016}{\natexlab{b}})}\BibitemShut {NoStop}%
\bibitem [{\citenamefont {Yashina}\ \emph {et~al.}(2008)\citenamefont
  {Yashina}, \citenamefont {P\"{u}ttner}, \citenamefont {Neudachina},
  \citenamefont {Zyubina}, \citenamefont {Shtanov},\ and\ \citenamefont
  {Poygin}}]{Yashina2008}%
  \BibitemOpen
  \bibfield  {author} {\bibinfo {author} {\bibfnamefont {L.~V.}\ \bibnamefont
  {Yashina}}, \bibinfo {author} {\bibfnamefont {R.}~\bibnamefont
  {P\"{u}ttner}}, \bibinfo {author} {\bibfnamefont {V.~S.}\ \bibnamefont
  {Neudachina}}, \bibinfo {author} {\bibfnamefont {T.~S.}\ \bibnamefont
  {Zyubina}}, \bibinfo {author} {\bibfnamefont {V.~I.}\ \bibnamefont
  {Shtanov}}, \ and\ \bibinfo {author} {\bibfnamefont {M.~V.}\ \bibnamefont
  {Poygin}},\ }\href {\doibase 10.1063/1.2912958} {\bibfield  {journal}
  {\bibinfo  {journal} {J. Appl. Phys.}\ }\textbf {\bibinfo {volume} {103}},\
  \bibinfo {pages} {094909} (\bibinfo {year} {2008})}\BibitemShut {NoStop}%
\bibitem [{\citenamefont {Sun}\ \emph {et~al.}(2006)\citenamefont {Sun},
  \citenamefont {Zhou},\ and\ \citenamefont {Ahuja}}]{Sun2006}%
  \BibitemOpen
  \bibfield  {author} {\bibinfo {author} {\bibfnamefont {Z.}~\bibnamefont
  {Sun}}, \bibinfo {author} {\bibfnamefont {J.}~\bibnamefont {Zhou}}, \ and\
  \bibinfo {author} {\bibfnamefont {R.}~\bibnamefont {Ahuja}},\ }\href
  {\doibase 10.1103/physrevlett.96.055507} {\bibfield  {journal} {\bibinfo
  {journal} {Phys. Rev. Lett.}\ }\textbf {\bibinfo {volume} {96}},\ \bibinfo
  {pages} {055507} (\bibinfo {year} {2006})}\BibitemShut {NoStop}%
\bibitem [{\citenamefont {Wang}\ and\ \citenamefont {Gedik}(2013)}]{Wang2013}%
  \BibitemOpen
  \bibfield  {author} {\bibinfo {author} {\bibfnamefont {Y.}~\bibnamefont
  {Wang}}\ and\ \bibinfo {author} {\bibfnamefont {N.}~\bibnamefont {Gedik}},\
  }\href {\doibase 10.1002/pssr.201206458} {\bibfield  {journal} {\bibinfo
  {journal} {Phys. Status Solidi-R}\ }\textbf {\bibinfo {volume} {7}},\
  \bibinfo {pages} {64} (\bibinfo {year} {2013})}\BibitemShut {NoStop}%
\bibitem [{\citenamefont {Scholz}\ \emph {et~al.}(2013)\citenamefont {Scholz},
  \citenamefont {S\'{a}nchez-Barriga}, \citenamefont {Braun}, \citenamefont
  {Marchenko}, \citenamefont {Varykhalov}, \citenamefont {Lindroos},
  \citenamefont {Wang}, \citenamefont {Lin}, \citenamefont {Bansil},
  \citenamefont {Min\'{a}r}, \citenamefont {Ebert}, \citenamefont {Volykhov},
  \citenamefont {Yashina},\ and\ \citenamefont {Rader}}]{Scholz2013}%
  \BibitemOpen
  \bibfield  {author} {\bibinfo {author} {\bibfnamefont {M.~R.}\ \bibnamefont
  {Scholz}}, \bibinfo {author} {\bibfnamefont {J.}~\bibnamefont
  {S\'{a}nchez-Barriga}}, \bibinfo {author} {\bibfnamefont {J.}~\bibnamefont
  {Braun}}, \bibinfo {author} {\bibfnamefont {D.}~\bibnamefont {Marchenko}},
  \bibinfo {author} {\bibfnamefont {A.}~\bibnamefont {Varykhalov}}, \bibinfo
  {author} {\bibfnamefont {M.}~\bibnamefont {Lindroos}}, \bibinfo {author}
  {\bibfnamefont {Y.~J.}\ \bibnamefont {Wang}}, \bibinfo {author}
  {\bibfnamefont {H.}~\bibnamefont {Lin}}, \bibinfo {author} {\bibfnamefont
  {A.}~\bibnamefont {Bansil}}, \bibinfo {author} {\bibfnamefont
  {J.}~\bibnamefont {Min\'{a}r}}, \bibinfo {author} {\bibfnamefont
  {H.}~\bibnamefont {Ebert}}, \bibinfo {author} {\bibfnamefont
  {A.}~\bibnamefont {Volykhov}}, \bibinfo {author} {\bibfnamefont {L.~V.}\
  \bibnamefont {Yashina}}, \ and\ \bibinfo {author} {\bibfnamefont
  {O.}~\bibnamefont {Rader}},\ }\href {\doibase 10.1103/PhysRevLett.110.216801}
  {\bibfield  {journal} {\bibinfo  {journal} {Phys. Rev. Lett.}\ }\textbf
  {\bibinfo {volume} {110}},\ \bibinfo {pages} {216801} (\bibinfo {year}
  {2013})}\BibitemShut {NoStop}%
\bibitem [{\citenamefont {Seibel}\ \emph {et~al.}(2015)\citenamefont {Seibel},
  \citenamefont {Maa{\ss}}, \citenamefont {Bentmann}, \citenamefont {Braun},
  \citenamefont {Sakamoto}, \citenamefont {Arita}, \citenamefont {Shimada},
  \citenamefont {Min\'{a}r}, \citenamefont {Ebert},\ and\ \citenamefont
  {Reinert}}]{Seibel2015}%
  \BibitemOpen
  \bibfield  {author} {\bibinfo {author} {\bibfnamefont {C.}~\bibnamefont
  {Seibel}}, \bibinfo {author} {\bibfnamefont {H.}~\bibnamefont {Maa{\ss}}},
  \bibinfo {author} {\bibfnamefont {H.}~\bibnamefont {Bentmann}}, \bibinfo
  {author} {\bibfnamefont {J.}~\bibnamefont {Braun}}, \bibinfo {author}
  {\bibfnamefont {K.}~\bibnamefont {Sakamoto}}, \bibinfo {author}
  {\bibfnamefont {M.}~\bibnamefont {Arita}}, \bibinfo {author} {\bibfnamefont
  {K.}~\bibnamefont {Shimada}}, \bibinfo {author} {\bibfnamefont
  {J.}~\bibnamefont {Min\'{a}r}}, \bibinfo {author} {\bibfnamefont
  {H.}~\bibnamefont {Ebert}}, \ and\ \bibinfo {author} {\bibfnamefont
  {F.}~\bibnamefont {Reinert}},\ }\href {\doibase 10.1016/j.elspec.2014.12.003}
  {\bibfield  {journal} {\bibinfo  {journal} {J. Electron Spectrosc.}\ }\textbf
  {\bibinfo {volume} {201}},\ \bibinfo {pages} {110} (\bibinfo {year}
  {2015})}\BibitemShut {NoStop}%
\bibitem [{\citenamefont {Kim}\ \emph {et~al.}(2010)\citenamefont {Kim},
  \citenamefont {Kim},\ and\ \citenamefont {Jhi}}]{Kim2010}%
  \BibitemOpen
  \bibfield  {author} {\bibinfo {author} {\bibfnamefont {J.}~\bibnamefont
  {Kim}}, \bibinfo {author} {\bibfnamefont {J.}~\bibnamefont {Kim}}, \ and\
  \bibinfo {author} {\bibfnamefont {S.-H.}\ \bibnamefont {Jhi}},\ }\href
  {\doibase 10.1103/physrevb.82.201312} {\bibfield  {journal} {\bibinfo
  {journal} {Phys. Rev. B}\ }\textbf {\bibinfo {volume} {82}},\ \bibinfo
  {pages} {201312} (\bibinfo {year} {2010})}\BibitemShut {NoStop}%
\bibitem [{\citenamefont {Sa}\ \emph {et~al.}(2011)\citenamefont {Sa},
  \citenamefont {Zhou}, \citenamefont {Song}, \citenamefont {Sun},\ and\
  \citenamefont {Ahuja}}]{Sa2011}%
  \BibitemOpen
  \bibfield  {author} {\bibinfo {author} {\bibfnamefont {B.}~\bibnamefont
  {Sa}}, \bibinfo {author} {\bibfnamefont {J.}~\bibnamefont {Zhou}}, \bibinfo
  {author} {\bibfnamefont {Z.}~\bibnamefont {Song}}, \bibinfo {author}
  {\bibfnamefont {Z.}~\bibnamefont {Sun}}, \ and\ \bibinfo {author}
  {\bibfnamefont {R.}~\bibnamefont {Ahuja}},\ }\href {\doibase
  10.1103/physrevb.84.085130} {\bibfield  {journal} {\bibinfo  {journal} {Phys.
  Rev. B}\ }\textbf {\bibinfo {volume} {84}},\ \bibinfo {pages} {085130}
  (\bibinfo {year} {2011})}\BibitemShut {NoStop}%
\bibitem [{\citenamefont {Kim}\ \emph {et~al.}(2012)\citenamefont {Kim},
  \citenamefont {Kim}, \citenamefont {Kim},\ and\ \citenamefont
  {Jhi}}]{Kim2012}%
  \BibitemOpen
  \bibfield  {author} {\bibinfo {author} {\bibfnamefont {J.}~\bibnamefont
  {Kim}}, \bibinfo {author} {\bibfnamefont {J.}~\bibnamefont {Kim}}, \bibinfo
  {author} {\bibfnamefont {K.-S.}\ \bibnamefont {Kim}}, \ and\ \bibinfo
  {author} {\bibfnamefont {S.-H.}\ \bibnamefont {Jhi}},\ }\href {\doibase
  10.1103/physrevlett.109.146601} {\bibfield  {journal} {\bibinfo  {journal}
  {Phys. Rev. Lett.}\ }\textbf {\bibinfo {volume} {109}},\ \bibinfo {pages}
  {146601} (\bibinfo {year} {2012})}\BibitemShut {NoStop}%
\bibitem [{\citenamefont {Sa}\ \emph {et~al.}(2012)\citenamefont {Sa},
  \citenamefont {Zhou}, \citenamefont {Sun},\ and\ \citenamefont
  {Ahuja}}]{Sa2012}%
  \BibitemOpen
  \bibfield  {author} {\bibinfo {author} {\bibfnamefont {B.}~\bibnamefont
  {Sa}}, \bibinfo {author} {\bibfnamefont {J.}~\bibnamefont {Zhou}}, \bibinfo
  {author} {\bibfnamefont {Z.}~\bibnamefont {Sun}}, \ and\ \bibinfo {author}
  {\bibfnamefont {R.}~\bibnamefont {Ahuja}},\ }\href {\doibase
  10.1209/0295-5075/97/27003} {\bibfield  {journal} {\bibinfo  {journal}
  {Europhys. Lett.}\ }\textbf {\bibinfo {volume} {97}},\ \bibinfo {pages}
  {27003} (\bibinfo {year} {2012})}\BibitemShut {NoStop}%
\bibitem [{\citenamefont {Silkin}\ \emph {et~al.}(2013)\citenamefont {Silkin},
  \citenamefont {Koroteev}, \citenamefont {Bihlmayer},\ and\ \citenamefont
  {Chulkov}}]{Silkin2013}%
  \BibitemOpen
  \bibfield  {author} {\bibinfo {author} {\bibfnamefont {I.~V.}\ \bibnamefont
  {Silkin}}, \bibinfo {author} {\bibfnamefont {Y.~M.}\ \bibnamefont
  {Koroteev}}, \bibinfo {author} {\bibfnamefont {G.}~\bibnamefont {Bihlmayer}},
  \ and\ \bibinfo {author} {\bibfnamefont {E.}~\bibnamefont {Chulkov}},\ }\href
  {\doibase 10.1016/j.apsusc.2012.09.017} {\bibfield  {journal} {\bibinfo
  {journal} {Appl. Surf. Sci.}\ }\textbf {\bibinfo {volume} {267}},\ \bibinfo
  {pages} {169} (\bibinfo {year} {2013})}\BibitemShut {NoStop}%
\bibitem [{\citenamefont {Pauly}\ \emph {et~al.}(2013)\citenamefont {Pauly},
  \citenamefont {Liebmann}, \citenamefont {Giussani}, \citenamefont {Kellner},
  \citenamefont {Just}, \citenamefont {S{\'{a}}nchez-Barriga}, \citenamefont
  {Rienks}, \citenamefont {Rader}, \citenamefont {Calarco}, \citenamefont
  {Bihlmayer},\ and\ \citenamefont {Morgenstern}}]{Pauly2013}%
  \BibitemOpen
  \bibfield  {author} {\bibinfo {author} {\bibfnamefont {C.}~\bibnamefont
  {Pauly}}, \bibinfo {author} {\bibfnamefont {M.}~\bibnamefont {Liebmann}},
  \bibinfo {author} {\bibfnamefont {A.}~\bibnamefont {Giussani}}, \bibinfo
  {author} {\bibfnamefont {J.}~\bibnamefont {Kellner}}, \bibinfo {author}
  {\bibfnamefont {S.}~\bibnamefont {Just}}, \bibinfo {author} {\bibfnamefont
  {J.}~\bibnamefont {S{\'{a}}nchez-Barriga}}, \bibinfo {author} {\bibfnamefont
  {E.}~\bibnamefont {Rienks}}, \bibinfo {author} {\bibfnamefont
  {O.}~\bibnamefont {Rader}}, \bibinfo {author} {\bibfnamefont
  {R.}~\bibnamefont {Calarco}}, \bibinfo {author} {\bibfnamefont
  {G.}~\bibnamefont {Bihlmayer}}, \ and\ \bibinfo {author} {\bibfnamefont
  {M.}~\bibnamefont {Morgenstern}},\ }\href {\doibase 10.1063/1.4847715}
  {\bibfield  {journal} {\bibinfo  {journal} {Appl. Phys. Lett.}\ }\textbf
  {\bibinfo {volume} {103}},\ \bibinfo {pages} {243109} (\bibinfo {year}
  {2013})}\BibitemShut {NoStop}%
\bibitem [{\citenamefont {Koirala}\ \emph {et~al.}(2015)\citenamefont
  {Koirala}, \citenamefont {Brahlek}, \citenamefont {Salehi}, \citenamefont
  {Wu}, \citenamefont {Dai}, \citenamefont {Waugh}, \citenamefont {Nummy},
  \citenamefont {Han}, \citenamefont {Moon}, \citenamefont {Zhu}, \citenamefont
  {Dessau}, \citenamefont {Wu}, \citenamefont {Armitage},\ and\ \citenamefont
  {Oh}}]{Oh2015}%
  \BibitemOpen
  \bibfield  {author} {\bibinfo {author} {\bibfnamefont {N.}~\bibnamefont
  {Koirala}}, \bibinfo {author} {\bibfnamefont {M.}~\bibnamefont {Brahlek}},
  \bibinfo {author} {\bibfnamefont {M.}~\bibnamefont {Salehi}}, \bibinfo
  {author} {\bibfnamefont {L.}~\bibnamefont {Wu}}, \bibinfo {author}
  {\bibfnamefont {J.}~\bibnamefont {Dai}}, \bibinfo {author} {\bibfnamefont
  {J.}~\bibnamefont {Waugh}}, \bibinfo {author} {\bibfnamefont
  {T.}~\bibnamefont {Nummy}}, \bibinfo {author} {\bibfnamefont {M.-G.}\
  \bibnamefont {Han}}, \bibinfo {author} {\bibfnamefont {J.}~\bibnamefont
  {Moon}}, \bibinfo {author} {\bibfnamefont {Y.}~\bibnamefont {Zhu}}, \bibinfo
  {author} {\bibfnamefont {D.}~\bibnamefont {Dessau}}, \bibinfo {author}
  {\bibfnamefont {W.}~\bibnamefont {Wu}}, \bibinfo {author} {\bibfnamefont
  {N.~P.}\ \bibnamefont {Armitage}}, \ and\ \bibinfo {author} {\bibfnamefont
  {S.}~\bibnamefont {Oh}},\ }\href {\doibase 10.1021/acs.nanolett.5b03770}
  {\bibfield  {journal} {\bibinfo  {journal} {Nano Lett.}\ }\textbf {\bibinfo
  {volume} {15}},\ \bibinfo {pages} {8245} (\bibinfo {year}
  {2015})}\BibitemShut {NoStop}%
\bibitem [{\citenamefont {Xu}\ \emph {et~al.}(2016)\citenamefont {Xu},
  \citenamefont {Miotkowski},\ and\ \citenamefont {Chen}}]{Xu2016}%
  \BibitemOpen
  \bibfield  {author} {\bibinfo {author} {\bibfnamefont {Y.}~\bibnamefont
  {Xu}}, \bibinfo {author} {\bibfnamefont {I.}~\bibnamefont {Miotkowski}}, \
  and\ \bibinfo {author} {\bibfnamefont {Y.~P.}\ \bibnamefont {Chen}},\ }\href
  {\doibase 10.1038/ncomms11434} {\bibfield  {journal} {\bibinfo  {journal}
  {Nat. Commun.}\ }\textbf {\bibinfo {volume} {7}},\ \bibinfo {pages} {11434}
  (\bibinfo {year} {2016})}\BibitemShut {NoStop}%
\bibitem [{\citenamefont {Medjanik}\ \emph {et~al.}(2017)\citenamefont
  {Medjanik}, \citenamefont {Fedchenko}, \citenamefont {Chernov}, \citenamefont
  {Kutnyakhov}, \citenamefont {Ellguth}, \citenamefont {Oelsner}, \citenamefont
  {Sch{\"o}nhense}, \citenamefont {Peixoto}, \citenamefont {Lutz},
  \citenamefont {Min}, \citenamefont {Reinert}, \citenamefont {D{\"a}ster},
  \citenamefont {Acremann}, \citenamefont {Viefhaus}, \citenamefont {Wurth},
  \citenamefont {Elmers},\ and\ \citenamefont {Sch{\"o}nhense}}]{Medjanik2017}%
  \BibitemOpen
  \bibfield  {author} {\bibinfo {author} {\bibfnamefont {K.}~\bibnamefont
  {Medjanik}}, \bibinfo {author} {\bibfnamefont {O.}~\bibnamefont {Fedchenko}},
  \bibinfo {author} {\bibfnamefont {S.}~\bibnamefont {Chernov}}, \bibinfo
  {author} {\bibfnamefont {D.}~\bibnamefont {Kutnyakhov}}, \bibinfo {author}
  {\bibfnamefont {M.}~\bibnamefont {Ellguth}}, \bibinfo {author} {\bibfnamefont
  {A.}~\bibnamefont {Oelsner}}, \bibinfo {author} {\bibfnamefont
  {B.}~\bibnamefont {Sch{\"o}nhense}}, \bibinfo {author} {\bibfnamefont
  {T.~R.~F.}\ \bibnamefont {Peixoto}}, \bibinfo {author} {\bibfnamefont
  {P.}~\bibnamefont {Lutz}}, \bibinfo {author} {\bibfnamefont {C.-H.}\
  \bibnamefont {Min}}, \bibinfo {author} {\bibfnamefont {F.}~\bibnamefont
  {Reinert}}, \bibinfo {author} {\bibfnamefont {S.}~\bibnamefont {D{\"a}ster}},
  \bibinfo {author} {\bibfnamefont {Y.}~\bibnamefont {Acremann}}, \bibinfo
  {author} {\bibfnamefont {J.}~\bibnamefont {Viefhaus}}, \bibinfo {author}
  {\bibfnamefont {W.}~\bibnamefont {Wurth}}, \bibinfo {author} {\bibfnamefont
  {H.~J.}\ \bibnamefont {Elmers}}, \ and\ \bibinfo {author} {\bibfnamefont
  {G.}~\bibnamefont {Sch{\"o}nhense}},\ }\href {\doibase 10.1038/nmat4875}
  {\bibfield  {journal} {\bibinfo  {journal} {Nat. Mater.}\ }\textbf {\bibinfo
  {volume} {16}},\ \bibinfo {pages} {615} (\bibinfo {year} {2017})}\BibitemShut
  {NoStop}%
\bibitem [{\citenamefont {H{\"u}fner}(2003)}]{Huefner}%
  \BibitemOpen
  \bibfield  {author} {\bibinfo {author} {\bibfnamefont {S.}~\bibnamefont
  {H{\"u}fner}},\ }\href@noop {} {\emph {\bibinfo {title} {Photoelectron
  Spectroscopy}}}\ (\bibinfo  {publisher} {Springer},\ \bibinfo {year}
  {2003})\BibitemShut {NoStop}%
\bibitem [{\citenamefont {Matzdorf}(1998)}]{Matsdorf1998}%
  \BibitemOpen
  \bibfield  {author} {\bibinfo {author} {\bibfnamefont {R.}~\bibnamefont
  {Matzdorf}},\ }\href {\doibase 10.1016/S0167-5729(97)00013-7} {\bibfield
  {journal} {\bibinfo  {journal} {Surf. Sci. Rep.}\ }\textbf {\bibinfo {volume}
  {30}},\ \bibinfo {pages} {153} (\bibinfo {year} {1998})}\BibitemShut
  {NoStop}%
\bibitem [{\citenamefont {Zhang}\ \emph {et~al.}(2011)\citenamefont {Zhang},
  \citenamefont {Chang}, \citenamefont {Zhang}, \citenamefont {Wen},
  \citenamefont {Feng}, \citenamefont {Li}, \citenamefont {Liu}, \citenamefont
  {He}, \citenamefont {Wang}, \citenamefont {Chen}, \citenamefont {Xue},
  \citenamefont {Ma},\ and\ \citenamefont {Wang}}]{Zhang2011}%
  \BibitemOpen
  \bibfield  {author} {\bibinfo {author} {\bibfnamefont {J.}~\bibnamefont
  {Zhang}}, \bibinfo {author} {\bibfnamefont {C.-Z.}\ \bibnamefont {Chang}},
  \bibinfo {author} {\bibfnamefont {Z.}~\bibnamefont {Zhang}}, \bibinfo
  {author} {\bibfnamefont {J.}~\bibnamefont {Wen}}, \bibinfo {author}
  {\bibfnamefont {X.}~\bibnamefont {Feng}}, \bibinfo {author} {\bibfnamefont
  {K.}~\bibnamefont {Li}}, \bibinfo {author} {\bibfnamefont {M.}~\bibnamefont
  {Liu}}, \bibinfo {author} {\bibfnamefont {K.}~\bibnamefont {He}}, \bibinfo
  {author} {\bibfnamefont {L.}~\bibnamefont {Wang}}, \bibinfo {author}
  {\bibfnamefont {X.}~\bibnamefont {Chen}}, \bibinfo {author} {\bibfnamefont
  {Q.-K.}\ \bibnamefont {Xue}}, \bibinfo {author} {\bibfnamefont
  {X.}~\bibnamefont {Ma}}, \ and\ \bibinfo {author} {\bibfnamefont
  {Y.}~\bibnamefont {Wang}},\ }\href {\doibase 10.1038/ncomms1588} {\bibfield
  {journal} {\bibinfo  {journal} {Nat. Commun.}\ }\textbf {\bibinfo {volume}
  {2}},\ \bibinfo {pages} {574} (\bibinfo {year} {2011})}\BibitemShut {NoStop}%
\bibitem [{\citenamefont {Nonaka}\ \emph {et~al.}(2000)\citenamefont {Nonaka},
  \citenamefont {Ohbayashi}, \citenamefont {Toriumi}, \citenamefont {Mori},\
  and\ \citenamefont {Hashimoto}}]{Nonaka2000}%
  \BibitemOpen
  \bibfield  {author} {\bibinfo {author} {\bibfnamefont {T.}~\bibnamefont
  {Nonaka}}, \bibinfo {author} {\bibfnamefont {G.}~\bibnamefont {Ohbayashi}},
  \bibinfo {author} {\bibfnamefont {Y.}~\bibnamefont {Toriumi}}, \bibinfo
  {author} {\bibfnamefont {Y.}~\bibnamefont {Mori}}, \ and\ \bibinfo {author}
  {\bibfnamefont {H.}~\bibnamefont {Hashimoto}},\ }\href {\doibase
  10.1016/S0040-6090(99)01090-1} {\bibfield  {journal} {\bibinfo  {journal}
  {Thin Solid Films}\ }\textbf {\bibinfo {volume} {370}},\ \bibinfo {pages}
  {258} (\bibinfo {year} {2000})}\BibitemShut {NoStop}%
\bibitem [{\citenamefont {Ando}\ \emph {et~al.}(1982)\citenamefont {Ando},
  \citenamefont {Fowler},\ and\ \citenamefont {Stern}}]{Ando}%
  \BibitemOpen
  \bibfield  {author} {\bibinfo {author} {\bibfnamefont {T.}~\bibnamefont
  {Ando}}, \bibinfo {author} {\bibfnamefont {A.~B.}\ \bibnamefont {Fowler}}, \
  and\ \bibinfo {author} {\bibfnamefont {F.}~\bibnamefont {Stern}},\ }\href
  {\doibase 10.1103/RevModPhys.54.437} {\bibfield  {journal} {\bibinfo
  {journal} {Rev. Mod. Phys.}\ }\textbf {\bibinfo {volume} {54}},\ \bibinfo
  {pages} {437} (\bibinfo {year} {1982})}\BibitemShut {NoStop}%
\bibitem [{\citenamefont {Edwards}\ \emph {et~al.}(2006)\citenamefont
  {Edwards}, \citenamefont {Pineda}, \citenamefont {Schultz}, \citenamefont
  {Martin}, \citenamefont {Thompson}, \citenamefont {Hjalmarson},\ and\
  \citenamefont {Umrigar}}]{Edwards2006}%
  \BibitemOpen
  \bibfield  {author} {\bibinfo {author} {\bibfnamefont {A.~H.}\ \bibnamefont
  {Edwards}}, \bibinfo {author} {\bibfnamefont {A.~C.}\ \bibnamefont {Pineda}},
  \bibinfo {author} {\bibfnamefont {P.~A.}\ \bibnamefont {Schultz}}, \bibinfo
  {author} {\bibfnamefont {M.~G.}\ \bibnamefont {Martin}}, \bibinfo {author}
  {\bibfnamefont {A.~P.}\ \bibnamefont {Thompson}}, \bibinfo {author}
  {\bibfnamefont {H.~P.}\ \bibnamefont {Hjalmarson}}, \ and\ \bibinfo {author}
  {\bibfnamefont {C.~J.}\ \bibnamefont {Umrigar}},\ }\href {\doibase
  10.1103/PhysRevB.73.045210} {\bibfield  {journal} {\bibinfo  {journal} {Phys.
  Rev. B}\ }\textbf {\bibinfo {volume} {73}},\ \bibinfo {pages} {045210}
  (\bibinfo {year} {2006})}\BibitemShut {NoStop}%
\bibitem [{\citenamefont {Wuttig}\ \emph {et~al.}(2006)\citenamefont {Wuttig},
  \citenamefont {L{\"u}sebrink}, \citenamefont {Wamwangi}, \citenamefont
  {We{\l}nic}, \citenamefont {Gille{\ss}en},\ and\ \citenamefont
  {Dronskowski}}]{Wuttig2007b}%
  \BibitemOpen
  \bibfield  {author} {\bibinfo {author} {\bibfnamefont {M.}~\bibnamefont
  {Wuttig}}, \bibinfo {author} {\bibfnamefont {D.}~\bibnamefont
  {L{\"u}sebrink}}, \bibinfo {author} {\bibfnamefont {D.}~\bibnamefont
  {Wamwangi}}, \bibinfo {author} {\bibfnamefont {W.}~\bibnamefont {We{\l}nic}},
  \bibinfo {author} {\bibfnamefont {M.}~\bibnamefont {Gille{\ss}en}}, \ and\
  \bibinfo {author} {\bibfnamefont {R.}~\bibnamefont {Dronskowski}},\ }\href
  {\doibase 10.1038/nmat1807} {\bibfield  {journal} {\bibinfo  {journal} {Nat.
  Mater.}\ }\textbf {\bibinfo {volume} {6}},\ \bibinfo {pages} {122} (\bibinfo
  {year} {2006})}\BibitemShut {NoStop}%
\bibitem [{\citenamefont {Lee}\ \emph {et~al.}(2005)\citenamefont {Lee},
  \citenamefont {Abelson}, \citenamefont {Bishop}, \citenamefont {Kang},
  \citenamefont {Cheong},\ and\ \citenamefont {Kim}}]{Lee2005}%
  \BibitemOpen
  \bibfield  {author} {\bibinfo {author} {\bibfnamefont {B.-S.}\ \bibnamefont
  {Lee}}, \bibinfo {author} {\bibfnamefont {J.~R.}\ \bibnamefont {Abelson}},
  \bibinfo {author} {\bibfnamefont {S.~G.}\ \bibnamefont {Bishop}}, \bibinfo
  {author} {\bibfnamefont {D.-H.}\ \bibnamefont {Kang}}, \bibinfo {author}
  {\bibfnamefont {B.-K.}\ \bibnamefont {Cheong}}, \ and\ \bibinfo {author}
  {\bibfnamefont {K.-B.}\ \bibnamefont {Kim}},\ }\href {\doibase
  10.1063/1.1884248} {\bibfield  {journal} {\bibinfo  {journal} {J. Appl.
  Phys.}\ }\textbf {\bibinfo {volume} {97}},\ \bibinfo {pages} {093509}
  (\bibinfo {year} {2005})}\BibitemShut {NoStop}%
\bibitem [{\citenamefont {Kellner}(2017)}]{Kellner}%
  \BibitemOpen
  \bibfield  {author} {\bibinfo {author} {\bibfnamefont {J.}~\bibnamefont
  {Kellner}},\ }\emph {\bibinfo {title} {A surface science based window to
  transport properties: The electronic structure of Te-based chalcogenides
  close to the Fermi level}},\ \href@noop {} {Ph.D. thesis},\ \bibinfo
  {school} {RWTH Aachen University} (\bibinfo {year} {2017})\BibitemShut
  {NoStop}%
\bibitem [{\citenamefont {Kellner}\ \emph {et~al.}(2017)\citenamefont
  {Kellner}, \citenamefont {Bihlmayer}, \citenamefont {Deringer}, \citenamefont
  {Liebmann}, \citenamefont {Pauly}, \citenamefont {Giussani}, \citenamefont
  {Boschker}, \citenamefont {Calarco}, \citenamefont {Dronskowski},\ and\
  \citenamefont {Morgenstern}}]{Kellner2017}%
  \BibitemOpen
  \bibfield  {author} {\bibinfo {author} {\bibfnamefont {J.}~\bibnamefont
  {Kellner}}, \bibinfo {author} {\bibfnamefont {G.}~\bibnamefont {Bihlmayer}},
  \bibinfo {author} {\bibfnamefont {V.~L.}\ \bibnamefont {Deringer}}, \bibinfo
  {author} {\bibfnamefont {M.}~\bibnamefont {Liebmann}}, \bibinfo {author}
  {\bibfnamefont {C.}~\bibnamefont {Pauly}}, \bibinfo {author} {\bibfnamefont
  {A.}~\bibnamefont {Giussani}}, \bibinfo {author} {\bibfnamefont {J.~E.}\
  \bibnamefont {Boschker}}, \bibinfo {author} {\bibfnamefont {R.}~\bibnamefont
  {Calarco}}, \bibinfo {author} {\bibfnamefont {R.}~\bibnamefont
  {Dronskowski}}, \ and\ \bibinfo {author} {\bibfnamefont {M.}~\bibnamefont
  {Morgenstern}},\ }\href {\doibase 10.1103/physrevb.96.245408} {\bibfield
  {journal} {\bibinfo  {journal} {Phys. Rev. B}\ }\textbf {\bibinfo {volume}
  {96}},\ \bibinfo {pages} {245408} (\bibinfo {year} {2017})}\BibitemShut
  {NoStop}%
\bibitem [{\citenamefont {Niesner}\ \emph {et~al.}(2012)\citenamefont
  {Niesner}, \citenamefont {Fauster}, \citenamefont {Eremeev}, \citenamefont
  {Menshchikova}, \citenamefont {Koroteev}, \citenamefont {Protogenov},
  \citenamefont {Chulkov}, \citenamefont {Tereshchenko}, \citenamefont {Kokh},
  \citenamefont {Alekperov}, \citenamefont {Nadjafov},\ and\ \citenamefont
  {Mamedov}}]{Niesner2012}%
  \BibitemOpen
  \bibfield  {author} {\bibinfo {author} {\bibfnamefont {D.}~\bibnamefont
  {Niesner}}, \bibinfo {author} {\bibfnamefont {T.}~\bibnamefont {Fauster}},
  \bibinfo {author} {\bibfnamefont {S.~V.}\ \bibnamefont {Eremeev}}, \bibinfo
  {author} {\bibfnamefont {T.~V.}\ \bibnamefont {Menshchikova}}, \bibinfo
  {author} {\bibfnamefont {Y.~M.}\ \bibnamefont {Koroteev}}, \bibinfo {author}
  {\bibfnamefont {A.~P.}\ \bibnamefont {Protogenov}}, \bibinfo {author}
  {\bibfnamefont {E.~V.}\ \bibnamefont {Chulkov}}, \bibinfo {author}
  {\bibfnamefont {O.~E.}\ \bibnamefont {Tereshchenko}}, \bibinfo {author}
  {\bibfnamefont {K.~A.}\ \bibnamefont {Kokh}}, \bibinfo {author}
  {\bibfnamefont {O.}~\bibnamefont {Alekperov}}, \bibinfo {author}
  {\bibfnamefont {A.}~\bibnamefont {Nadjafov}}, \ and\ \bibinfo {author}
  {\bibfnamefont {N.}~\bibnamefont {Mamedov}},\ }\href {\doibase
  10.1103/PhysRevB.86.205403} {\bibfield  {journal} {\bibinfo  {journal} {Phys.
  Rev. B}\ }\textbf {\bibinfo {volume} {86}},\ \bibinfo {pages} {205403}
  (\bibinfo {year} {2012})}\BibitemShut {NoStop}%
\bibitem [{\citenamefont {Crepaldi}\ \emph {et~al.}(2014)\citenamefont
  {Crepaldi}, \citenamefont {Cilento}, \citenamefont {Zacchigna}, \citenamefont
  {Zonno}, \citenamefont {Johannsen}, \citenamefont {Tournier-Colletta},
  \citenamefont {Moreschini}, \citenamefont {Vobornik}, \citenamefont
  {Bondino}, \citenamefont {Magnano}, \citenamefont {Berger}, \citenamefont
  {Magrez}, \citenamefont {Bugnon}, \citenamefont {Autes}, \citenamefont
  {Yazyev}, \citenamefont {Grioni},\ and\ \citenamefont
  {Parmigiani}}]{Crepaldi2014}%
  \BibitemOpen
  \bibfield  {author} {\bibinfo {author} {\bibfnamefont {A.}~\bibnamefont
  {Crepaldi}}, \bibinfo {author} {\bibfnamefont {F.}~\bibnamefont {Cilento}},
  \bibinfo {author} {\bibfnamefont {M.}~\bibnamefont {Zacchigna}}, \bibinfo
  {author} {\bibfnamefont {M.}~\bibnamefont {Zonno}}, \bibinfo {author}
  {\bibfnamefont {J.~C.}\ \bibnamefont {Johannsen}}, \bibinfo {author}
  {\bibfnamefont {C.}~\bibnamefont {Tournier-Colletta}}, \bibinfo {author}
  {\bibfnamefont {L.}~\bibnamefont {Moreschini}}, \bibinfo {author}
  {\bibfnamefont {I.}~\bibnamefont {Vobornik}}, \bibinfo {author}
  {\bibfnamefont {F.}~\bibnamefont {Bondino}}, \bibinfo {author} {\bibfnamefont
  {E.}~\bibnamefont {Magnano}}, \bibinfo {author} {\bibfnamefont
  {H.}~\bibnamefont {Berger}}, \bibinfo {author} {\bibfnamefont
  {A.}~\bibnamefont {Magrez}}, \bibinfo {author} {\bibfnamefont
  {P.}~\bibnamefont {Bugnon}}, \bibinfo {author} {\bibfnamefont
  {G.}~\bibnamefont {Autes}}, \bibinfo {author} {\bibfnamefont {O.~V.}\
  \bibnamefont {Yazyev}}, \bibinfo {author} {\bibfnamefont {M.}~\bibnamefont
  {Grioni}}, \ and\ \bibinfo {author} {\bibfnamefont {F.}~\bibnamefont
  {Parmigiani}},\ }\href {\doibase 10.1103/PhysRevB.89.125408} {\bibfield
  {journal} {\bibinfo  {journal} {Phys. Rev. B}\ }\textbf {\bibinfo {volume}
  {89}},\ \bibinfo {pages} {125408} (\bibinfo {year} {2014})}\BibitemShut
  {NoStop}%
\bibitem [{\citenamefont {Hasan}\ and\ \citenamefont {Kane}(2010)}]{Hasan2008}%
  \BibitemOpen
  \bibfield  {author} {\bibinfo {author} {\bibfnamefont {M.~Z.}\ \bibnamefont
  {Hasan}}\ and\ \bibinfo {author} {\bibfnamefont {C.~L.}\ \bibnamefont
  {Kane}},\ }\href {\doibase 10.1103/revmodphys.82.3045} {\bibfield  {journal}
  {\bibinfo  {journal} {Rev. Mod. Phys.}\ }\textbf {\bibinfo {volume} {82}},\
  \bibinfo {pages} {3045} (\bibinfo {year} {2010})}\BibitemShut {NoStop}%
\bibitem [{\citenamefont {Qi}\ and\ \citenamefont {Zhang}(2011)}]{Qi2011}%
  \BibitemOpen
  \bibfield  {author} {\bibinfo {author} {\bibfnamefont {X.-L.}\ \bibnamefont
  {Qi}}\ and\ \bibinfo {author} {\bibfnamefont {S.-C.}\ \bibnamefont {Zhang}},\
  }\href {\doibase 10.1103/revmodphys.83.1057} {\bibfield  {journal} {\bibinfo
  {journal} {Rev. Mod. Phys.}\ }\textbf {\bibinfo {volume} {83}},\ \bibinfo
  {pages} {1057} (\bibinfo {year} {2011})}\BibitemShut {NoStop}%
\bibitem [{\citenamefont {Ando}(2013)}]{Ando2013}%
  \BibitemOpen
  \bibfield  {author} {\bibinfo {author} {\bibfnamefont {Y.}~\bibnamefont
  {Ando}},\ }\href {\doibase 10.7566/JPSJ.82.102001} {\bibfield  {journal}
  {\bibinfo  {journal} {J. Phys. Soc. Jpn.}\ }\textbf {\bibinfo {volume}
  {82}},\ \bibinfo {pages} {102001} (\bibinfo {year} {2013})}\BibitemShut
  {NoStop}%
\bibitem [{\citenamefont {Pauly}\ \emph {et~al.}(2012)\citenamefont {Pauly},
  \citenamefont {Bihlmayer}, \citenamefont {Liebmann}, \citenamefont {Grob},
  \citenamefont {Georgi}, \citenamefont {Subramaniam}, \citenamefont {Scholz},
  \citenamefont {S{\'{a}}nchez-Barriga}, \citenamefont {Varykhalov},
  \citenamefont {Bl{\"u}gel}, \citenamefont {Rader},\ and\ \citenamefont
  {Morgenstern}}]{Pauly2012}%
  \BibitemOpen
  \bibfield  {author} {\bibinfo {author} {\bibfnamefont {C.}~\bibnamefont
  {Pauly}}, \bibinfo {author} {\bibfnamefont {G.}~\bibnamefont {Bihlmayer}},
  \bibinfo {author} {\bibfnamefont {M.}~\bibnamefont {Liebmann}}, \bibinfo
  {author} {\bibfnamefont {M.}~\bibnamefont {Grob}}, \bibinfo {author}
  {\bibfnamefont {A.}~\bibnamefont {Georgi}}, \bibinfo {author} {\bibfnamefont
  {D.}~\bibnamefont {Subramaniam}}, \bibinfo {author} {\bibfnamefont {M.~R.}\
  \bibnamefont {Scholz}}, \bibinfo {author} {\bibfnamefont {J.}~\bibnamefont
  {S{\'{a}}nchez-Barriga}}, \bibinfo {author} {\bibfnamefont {A.}~\bibnamefont
  {Varykhalov}}, \bibinfo {author} {\bibfnamefont {S.}~\bibnamefont
  {Bl{\"u}gel}}, \bibinfo {author} {\bibfnamefont {O.}~\bibnamefont {Rader}}, \
  and\ \bibinfo {author} {\bibfnamefont {M.}~\bibnamefont {Morgenstern}},\
  }\href {\doibase 10.1103/physrevb.86.235106} {\bibfield  {journal} {\bibinfo
  {journal} {Phys. Rev. B}\ }\textbf {\bibinfo {volume} {86}},\ \bibinfo
  {pages} {235106} (\bibinfo {year} {2012})}\BibitemShut {NoStop}%
\bibitem [{\citenamefont {Thomann}\ \emph {et~al.}(1999)\citenamefont
  {Thomann}, \citenamefont {Shumay}, \citenamefont {Weinelt},\ and\
  \citenamefont {Fauster}}]{Thomann1999}%
  \BibitemOpen
  \bibfield  {author} {\bibinfo {author} {\bibfnamefont {U.}~\bibnamefont
  {Thomann}}, \bibinfo {author} {\bibfnamefont {I.~L.}\ \bibnamefont {Shumay}},
  \bibinfo {author} {\bibfnamefont {M.}~\bibnamefont {Weinelt}}, \ and\
  \bibinfo {author} {\bibfnamefont {T.}~\bibnamefont {Fauster}},\ }\href
  {\doibase 10.1007/s003400050661} {\bibfield  {journal} {\bibinfo  {journal}
  {Appl. Phys. B}\ }\textbf {\bibinfo {volume} {68}},\ \bibinfo {pages} {531}
  (\bibinfo {year} {1999})}\BibitemShut {NoStop}%
\bibitem [{\citenamefont {Rieger}\ \emph {et~al.}(1983)\citenamefont {Rieger},
  \citenamefont {Schnell}, \citenamefont {Steinmann},\ and\ \citenamefont
  {Saile}}]{Rieger1983}%
  \BibitemOpen
  \bibfield  {author} {\bibinfo {author} {\bibfnamefont {D.}~\bibnamefont
  {Rieger}}, \bibinfo {author} {\bibfnamefont {R.~D.}\ \bibnamefont {Schnell}},
  \bibinfo {author} {\bibfnamefont {W.}~\bibnamefont {Steinmann}}, \ and\
  \bibinfo {author} {\bibfnamefont {V.}~\bibnamefont {Saile}},\ }\href
  {\doibase 10.1016/0167-5087(83)91220-6} {\bibfield  {journal} {\bibinfo
  {journal} {Nucl. Instrum. Methods}\ }\textbf {\bibinfo {volume} {208}},\
  \bibinfo {pages} {777} (\bibinfo {year} {1983})}\BibitemShut {NoStop}%
\bibitem [{\citenamefont {Zhang}\ \emph {et~al.}(2010)\citenamefont {Zhang},
  \citenamefont {Pan}, \citenamefont {Foo}, \citenamefont {Fang}, \citenamefont
  {Yeo}, \citenamefont {Zhao}, \citenamefont {Shi},\ and\ \citenamefont
  {Chong}}]{Zhang2010}%
  \BibitemOpen
  \bibfield  {author} {\bibinfo {author} {\bibfnamefont {Z.}~\bibnamefont
  {Zhang}}, \bibinfo {author} {\bibfnamefont {J.}~\bibnamefont {Pan}}, \bibinfo
  {author} {\bibfnamefont {Y.~L.}\ \bibnamefont {Foo}}, \bibinfo {author}
  {\bibfnamefont {L.~W.-W.}\ \bibnamefont {Fang}}, \bibinfo {author}
  {\bibfnamefont {Y.-C.}\ \bibnamefont {Yeo}}, \bibinfo {author} {\bibfnamefont
  {R.}~\bibnamefont {Zhao}}, \bibinfo {author} {\bibfnamefont {L.}~\bibnamefont
  {Shi}}, \ and\ \bibinfo {author} {\bibfnamefont {T.-C.}\ \bibnamefont
  {Chong}},\ }\href {\doibase 10.1016/j.apusc.2010.06.039} {\bibfield
  {journal} {\bibinfo  {journal} {Appl. Surf. Sci.}\ }\textbf {\bibinfo
  {volume} {256}},\ \bibinfo {pages} {7696} (\bibinfo {year}
  {2010})}\BibitemShut {NoStop}%
\bibitem [{\citenamefont {Gourvest}\ \emph {et~al.}(2012)\citenamefont
  {Gourvest}, \citenamefont {Pelissier}, \citenamefont {Vallee}, \citenamefont
  {Roule}, \citenamefont {Lhoutis},\ and\ \citenamefont
  {Maitrejean}}]{Gourvest2012}%
  \BibitemOpen
  \bibfield  {author} {\bibinfo {author} {\bibfnamefont {E.}~\bibnamefont
  {Gourvest}}, \bibinfo {author} {\bibfnamefont {B.}~\bibnamefont {Pelissier}},
  \bibinfo {author} {\bibfnamefont {C.}~\bibnamefont {Vallee}}, \bibinfo
  {author} {\bibfnamefont {A.}~\bibnamefont {Roule}}, \bibinfo {author}
  {\bibfnamefont {S.}~\bibnamefont {Lhoutis}}, \ and\ \bibinfo {author}
  {\bibfnamefont {S.}~\bibnamefont {Maitrejean}},\ }\href {\doibase
  10.1149/2.027204jes} {\bibfield  {journal} {\bibinfo  {journal} {J.
  Electrochem. Soc.}\ }\textbf {\bibinfo {volume} {159}},\ \bibinfo {pages}
  {H373} (\bibinfo {year} {2012})}\BibitemShut {NoStop}%
\bibitem [{\citenamefont {Deringer}\ \emph {et~al.}(2012)\citenamefont
  {Deringer}, \citenamefont {Lumeij},\ and\ \citenamefont
  {Dronskowski}}]{Deringer}%
  \BibitemOpen
  \bibfield  {author} {\bibinfo {author} {\bibfnamefont {V.~L.}\ \bibnamefont
  {Deringer}}, \bibinfo {author} {\bibfnamefont {M.}~\bibnamefont {Lumeij}}, \
  and\ \bibinfo {author} {\bibfnamefont {R.}~\bibnamefont {Dronskowski}},\
  }\href {\doibase 10.1021/jp304455z} {\bibfield  {journal} {\bibinfo
  {journal} {J. Phys. Chem. C}\ }\textbf {\bibinfo {volume} {116}},\ \bibinfo
  {pages} {15801} (\bibinfo {year} {2012})}\BibitemShut {NoStop}%
\bibitem [{\citenamefont {Ibach}\ and\ \citenamefont {L{\"u}th}(2009)}]{Ibach}%
  \BibitemOpen
  \bibfield  {author} {\bibinfo {author} {\bibfnamefont {H.}~\bibnamefont
  {Ibach}}\ and\ \bibinfo {author} {\bibfnamefont {H.}~\bibnamefont
  {L{\"u}th}},\ }\href@noop {} {\emph {\bibinfo {title} {Solid-State
  Physics}}}\ (\bibinfo  {publisher} {Springer},\ \bibinfo {year}
  {2009})\BibitemShut {NoStop}%
\bibitem [{\citenamefont {Kramer}\ and\ \citenamefont
  {MacKinnon}(1993)}]{Kramer1993}%
  \BibitemOpen
  \bibfield  {author} {\bibinfo {author} {\bibfnamefont {B.}~\bibnamefont
  {Kramer}}\ and\ \bibinfo {author} {\bibfnamefont {A.}~\bibnamefont
  {MacKinnon}},\ }\href {\doibase 10.1088/0034-4885/56/12/001} {\bibfield
  {journal} {\bibinfo  {journal} {Rep. Prog. Phys.}\ }\textbf {\bibinfo
  {volume} {56}},\ \bibinfo {pages} {1469} (\bibinfo {year}
  {1993})}\BibitemShut {NoStop}%
\bibitem [{\citenamefont {Evers}\ and\ \citenamefont
  {Mirlin}(2008)}]{Evers2008}%
  \BibitemOpen
  \bibfield  {author} {\bibinfo {author} {\bibfnamefont {F.}~\bibnamefont
  {Evers}}\ and\ \bibinfo {author} {\bibfnamefont {A.~D.}\ \bibnamefont
  {Mirlin}},\ }\href {\doibase 10.1103/RevModPhys.80.1355} {\bibfield
  {journal} {\bibinfo  {journal} {Rev. Mod. Phys.}\ }\textbf {\bibinfo {volume}
  {80}},\ \bibinfo {pages} {1355} (\bibinfo {year} {2008})}\BibitemShut
  {NoStop}%
\bibitem [{\citenamefont {Perdew}\ \emph {et~al.}(1996)\citenamefont {Perdew},
  \citenamefont {Burke},\ and\ \citenamefont {Ernzerhof}}]{Perdew1996}%
  \BibitemOpen
  \bibfield  {author} {\bibinfo {author} {\bibfnamefont {J.~P.}\ \bibnamefont
  {Perdew}}, \bibinfo {author} {\bibfnamefont {K.}~\bibnamefont {Burke}}, \
  and\ \bibinfo {author} {\bibfnamefont {M.}~\bibnamefont {Ernzerhof}},\ }\href
  {\doibase 10.1103/physrevlett.77.3865} {\bibfield  {journal} {\bibinfo
  {journal} {Phys. Rev. Lett.}\ }\textbf {\bibinfo {volume} {77}},\ \bibinfo
  {pages} {3865} (\bibinfo {year} {1996})}\BibitemShut {NoStop}%
\bibitem [{\citenamefont {Wimmer}\ \emph {et~al.}(1981)\citenamefont {Wimmer},
  \citenamefont {Krakauer}, \citenamefont {Weinert},\ and\ \citenamefont
  {Freeman}}]{Wimmer1981}%
  \BibitemOpen
  \bibfield  {author} {\bibinfo {author} {\bibfnamefont {E.}~\bibnamefont
  {Wimmer}}, \bibinfo {author} {\bibfnamefont {H.}~\bibnamefont {Krakauer}},
  \bibinfo {author} {\bibfnamefont {M.}~\bibnamefont {Weinert}}, \ and\
  \bibinfo {author} {\bibfnamefont {A.~J.}\ \bibnamefont {Freeman}},\ }\href
  {\doibase 10.1103/PhysRevB.24.864} {\bibfield  {journal} {\bibinfo  {journal}
  {Phys. Rev. B}\ }\textbf {\bibinfo {volume} {24}},\ \bibinfo {pages} {864}
  (\bibinfo {year} {1981})}\BibitemShut {NoStop}%
\end{thebibliography}

%

\end{document}